\documentclass[journal=jacsat,manuscript=article,layout=traditional]{achemso}

\usepackage[version=3]{mhchem}
\usepackage{amssymb}
\usepackage{tikz}
\usepackage{booktabs}
\usepackage{tabularx}
\usepackage{pgfplots}
\usepackage{amsmath}
\pgfplotsset{compat=1.9}
\usetikzlibrary{shapes,arrows,positioning}

\SectionNumbersOn
\author{Alicia van Hees}
\affiliation[UU]{Department of Chemistry-\AA{}ngstr\"om Laboratory, Uppsala
  University, L\"agerhyddsv\"agen 1, BOX 538, 75121 Uppsala, Sweden}
\author{Zhan-Yun Zhang}
\affiliation[UU]{Department of Chemistry-\AA{}ngstr\"om Laboratory, Uppsala
  University, L\"agerhyddsv\"agen 1, BOX 538, 75121 Uppsala, Sweden}
\author{Aishwarya Sudhama}
\affiliation[UU]{Department of Chemistry-\AA{}ngstr\"om Laboratory, Uppsala
  University, L\"agerhyddsv\"agen 1, BOX 538, 75121 Uppsala, Sweden}
\author{Chao Zhang}
\affiliation[UU]{Department of Chemistry-\AA{}ngstr\"om Laboratory, Uppsala
  University, L\"agerhyddsv\"agen 1, BOX 538, 75121 Uppsala, Sweden}
\alsoaffiliation{Wallenberg Initiative Materials Science for Sustainability, Uppsala University, 75121 Uppsala, Sweden}
\email{chao.zhang@kemi.uu.se}

\title{Molecular Modelling of Aqueous Batteries}

\begin{document}


\begin{abstract}
Aqueous batteries play an increasingly important role for the development of sustainable and safety-prioritised energy storage solutions. Compared to conventional lithium-ion batteries, the cell chemistry in aqueous batteries share many common features with those of electrolyzer and pseudo-capacitor systems because of the involvement of aqueous electrolyte and proton activity. This imposes the needs for a better understanding of the corresponding ion solvation, intercalation and electron transfer processes at atomistic scale. Therefore, this chapter provides an up-to-date overview of molecular modelling techniques and their applications in aqueous batteries. In particular, we emphasize on the dynamical and reactive description of aqueous battery systems brought in by density functional theory-based molecular dynamics simulation (DFTMD) and its machine-learning (ML) accelerated counterpart. Moreover, we also cover the recent advancement of generative artificial intelligence (AI) in molecular and materials design of aqueous batteries. Case studies presented here include popular aqueous battery systems, such as water-in-salt electrolytes, proton-coupled cathode materials, Zn-ion batteries as well as organic redox flow batteries.
 
\end{abstract}

\section{Introduction}

Electrochemical energy storage systems are indispensable components for building a sustainable and fossil-free society with infrastructures such as electric vehicles and energy grids. In particular, rechargeable batteries have attracted an ever-increasing attention in research going from the materials chemistry to the cell manufacturing, which was highlighted by the 2019 Nobel Prize in Chemistry. In this ongoing endeavours, aqueous batteries are a promising direction thanks to their high ion conductivity (and power density), low risk of thermal runaway and cheap manufacturing cost~\cite{Liu2019, 2020.Borodin, 10.1039/d1ee02021h, 2021.Lijae, 10.1002/anie.202200598, 2023.Zhangxo, 10.1002/adma.202300053, 2023.Ahnzy, 10.1002/smtd.202300699}. On the other hands, the scale-up of aqueous battery systems faces the challenges of the intrinsic narrow electrochemical stability window of water (low energy density),  unstable solid-electrolyte interphase as well as swelling (due to the water splitting). 

In order to overcome these challenges, to disentangle the complexity in aqueous battery systems and to advance the field through fundamental understanding in addition to engineering efforts, a physical chemistry approach is clearly needed. Good examples includes charting highly concentrated water-in-salt electrolytes in Li-ion batteries that can stand up to 3V high voltage which is way beyond their thought-to-be electrochemical stability~\cite{Suo2015WiSE} and exploiting the idea of the out sphere electron transfer in Zn-ion batteries to mitigate the dendrite growth~\cite{2024.Zhanghjj}. Thus, to find out how interfacial confinement, ion concentration, and applied potential interplay with each other and reshape the structural, transport, and electrical properties of electrolyte materials and associated interfaces, microscopic information and its connections to macroscopic properties are highly desirable. 

This crucial information can be very difficult to extract from the top-down approach commonly used in experiments. Therefore, molecular modelling built in a bottom-up fashion becomes complementary and indispensable~\cite{2023acfe9b}. Based on the principles of quantum mechanics and statistical mechanics, density functional theory (DFT) calculations and DFT-based molecular dynamics (DFTMD) simulations are eminently suitable techniques for modeling electrolytes and electrolyte interfaces where the distinction between reactive solutes and solvent has all but disappeared. Nevertheless, many electrochemical properties of bulk electrolytes and electrolyte interfaces and the corresponding processes, e.g. fully converged ionic conductivities and charging dynamics, cannot possibly be described with a few hundred atoms and tens of picoseconds with DFTMD.  Therefore, data-driven approaches such as machine learning force field (MLFF) have emerged as the game changer, which can describe reactive systems with quantum mechanical accuracy but only for a fraction of computational costs~\cite{shao2021modelling}. 

As compared to other branches of computational (theoretical) chemistry, the boundary condition matters in computational electrochemistry as much as it does in potentiostatic, galvanostatic, or coulostatic measurements in electrochemical experiments. This means conceptual gaps can persist between simulation and experiment even if we can find the “best Hamiltonian” and compute it cheaply over the desired length and time scales. A notable example is on the investigation the true transference number of polymer electrolytes~\cite{2024.Shao}, where a unified theoretical framework can be developed to take this point into account, which otherwise led to significant confusion and controversy in the field. 

Therefore, in this chapter, we will first introduce the essence of molecular modelling techniques including the generative artificial intelligence (AI) which shows a high potential to revolutionize the field of molecular and materials design. Then, it is followed by case studies and discussions of popular aqueous battery systems including water-in-salt electrolytes, proton-coupled cathode materials, Zn-ion batteries, and organic redox flow batteries. When selecting these examples, we emphasized on studies where molecular modelling played a substantial rather than superficial role. This hopefully can also provide ideas about how to approach experimental battery research from theory and simulation (and vice versa). 

\section{Molecular modelling methods}

\subsection{Electronic structure calculations and simulations}

Density functional theory (DFT) and density functional theory based molecular dynamics (DFTMD) are the working horses for the molecular modelling of battery materials. They can in principle provide a consistent description of the structural, dynamical, energetic and electrical properties of condensed phase systems such as electrode-electrolyte interfaces and electrolyte materials. In addition, they often provide the reference calculations for parameterizing molecular mechanics or machine learning force fields. That is why we focus on them in this session rather than their more accurate but computationally expensive counterparts -- wave function-based methods.

\subsubsection{Density functional theory}
On a fundamental level, the properties of molecular systems are determined by the quantum mechanical nature of their constituent particles. 
While atomic nuclei are large enough for classical approximations to hold in most cases, electrons require a quantum mechanical treatment. Computational methods that seek to solve the electronic structure of a molecule or material thus build on some way of approximating solutions to the many-body electronic wave function.  

\begin{figure}
    \centering
    \includegraphics{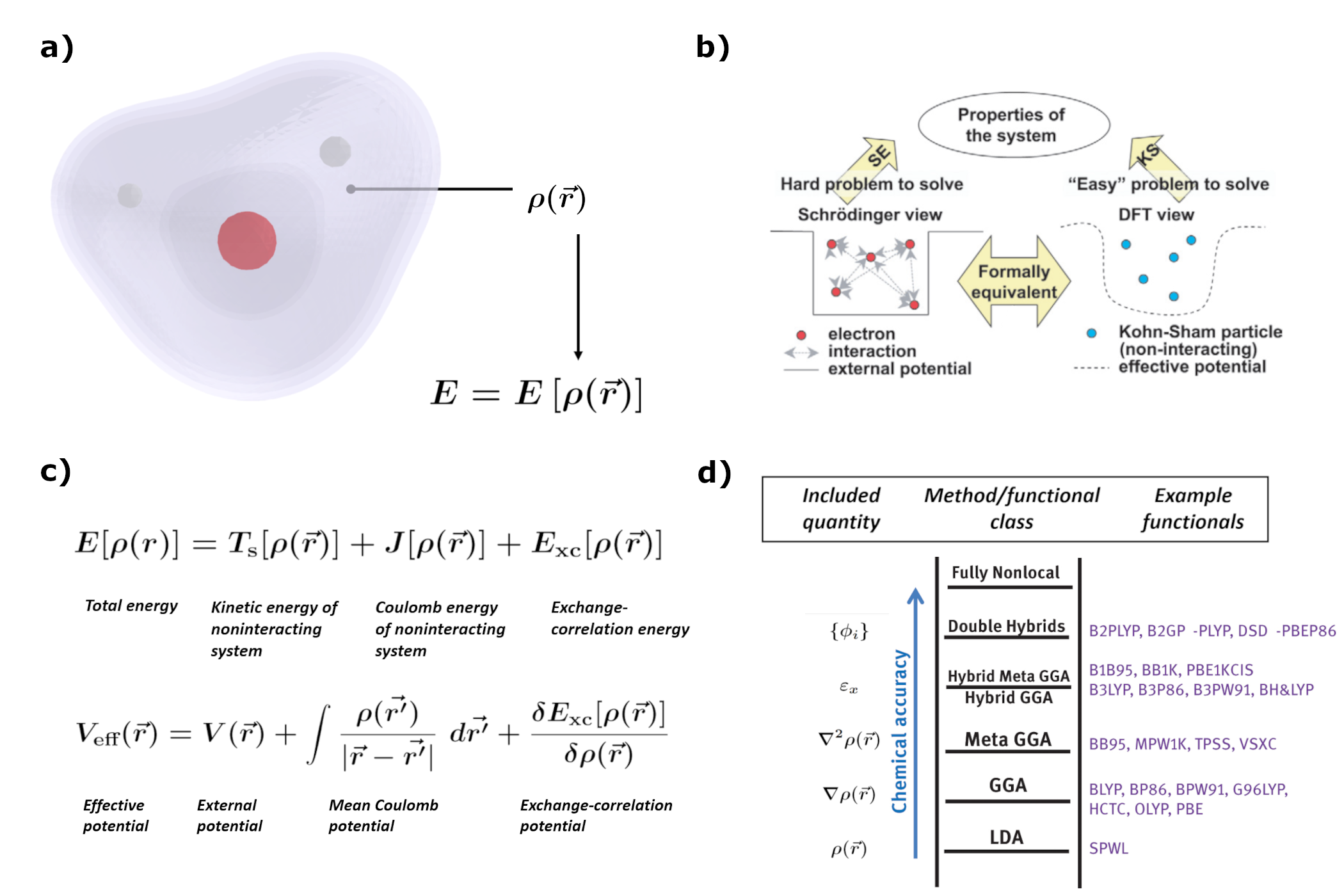}
    \caption{(a) Within density function theory, the energy $E$ is described as a functional of the electron density $\rho$, which is a function of position in space ($\Vec{r}$). (b) Graphic showing the difference between Kohn-Sham DFT (KS) and the full Schrödinger equation (SE). Adapted with permission from  \cite{Mattsson2005} (Copyright 2005 IOP Publishing Ltd). (c) The energy functional of electron density as partitioned within Kohn-Sham DFT, and the effective potential of the non-interacting electrons within the Kohn-Shan formalism. (d) The so-called ``Jacobs ladder", showing how more intricate functionals lead to estimations closer to the full Schrödinger equation. From the local density approximations LDA, which includes only the local value of the electron density $\rho(\Vec{r})$, the functionals get more intricate as dependence on the density gradient $\nabla \rho(\Vec{r})$ and Laplacian $\nabla^2 \rho(\Vec{r})$ are included. In hybrid functionals, exact HF exchange is added, relating the system description more closely to the occupied wave function orbital energies $\epsilon_x$, and finally the double hybrid orbitals add a dependence also on unoccupied orbitals $\phi_i$ though perturbation theory. Adapted with permission from \cite{Palafox2019} (Copyright 2019 Walter de Gruyter).}
    \label{fig_dft}
\end{figure}

Besides working directly with the wave function, powerful computational frameworks and tools have also been developed within density functional theory (DFT). Hohenberg and Kohn \cite{Hohenberg_Kohn} showed that, if the electronic density of a system and the position of its nuclei is known, all ground state properties can in principle be found, given that a functional is known that maps every density to its unique ground state energy. The reformulation of quantum mechanics with electron density, which is a 3 dimensional function, allows to circumvent the challenge to describing the high dimensional (wave) function of the whole system (Figure \ref{fig_dft}a).  

 Despite of its great theoretical promise, this universal functional in DFT is not known in practice. The whole field of DFT is built around functionals that approximate the ground state energy of a given electron density, and these different approximations have led to the zoo of DFT functionals in use today. Due to its high efficiency and usefulness, most of these functionals are rooted in the Kohn-Sham formalism first presented 1965 \cite{Kohn_Sham}. Here, the density of the system is described as a composition (the so-called Slater determinant) of non-interacting one-electron wave functions that has come to be known as Kohn-Sham orbitals. Interactions between electrons are reformulated in the external potential in which these Kohn-Sham orbitals are solved for (Figure \ref{fig_dft}b, c). The role of the exchange-correlation functional in this external potential is to include quantum exchange and correlation of the electrons. Accordingly, different exchange-correlation functionals have been developed over the years. Thorough mathematical descriptions of these functional classes can be found in many textbooks, for example the one by Koch and Holthausen \cite{Chemists_DFT}. Here, we will just provide a brief overview of them. See also Figure \ref{fig_dft}d for a hierarchical outline of different functional classes.

One early class of functionals are based on the local density approximation, and are thus known as LDA functionals. The exchange-correlation energy contribution from every infinitesimal volume of density is treated as the homogeneous electron gas in which the exchange energy can be calculated exactly, and highly accurate estimations of the correlation energy can be obtained from quantum Monte Carlo simulations. Later, ways to improve the accuracy by including also the local value of the density gradient in the functional were introduced. The most successful of these new classes of functionals, known as the generalised gradient approximation (GGA) functionals, performs much more satisfactory for chemical systems than LDA functionals \cite{Chemists_DFT}. There are also functionals whose mathematical expressions include dependence on the second order derivatives, often in the form of the kinetic energy density. These functionals are often referred to as meta-GGA functionals. Within the wave function based Hartree-Fock theory, an analytical expression for the exchange energy of non-interacting Kohn-Sham orbitals exists. Including a fraction of this contribution to the exchange correlation energy created the family of so-called hybrid functionals. In addition, DFT calculations can also be improved upon by applying empirical or semi-empirical corrections. Common examples of such corrections are the practice of correction for dispersion interaction, for example following the example of Grimme and co-workers \cite{Grimme}, and the U correction for strong electronic correlation \cite{Rousseau2020}. 

 Before ending this section, it is worth noting that there are theoretical and practical limitations of DFT. As a ground state theory, only the energy of the highest occupied molecular orbital (HOMO) has a clear physical meaning in DFT, which corresponds to the negative of the first ionization potential. Nevertheless, Kohn-Sham orbitals have been shown to have merit for providing a qualitative picture even if this was not promised by the theory \cite{Marzari2021}. Then, a known practical limitation of DFT, is the accurate estimation of band gaps. Similarly, many GGA functional failed the systems with high degrees of electron correlation, such as reducible metal oxides. In this regard, hybrid and other newly developed meta-GGA  functionals perform often better\cite{Borlido2019}. 

\subsubsection{Density functional theory-based molecular dynamics}

As previously mentioned, DFT is often used to calculate the equilibrium energy of molecular systems through geometry optimisation and single-point energy calculations. However, high performance computers (HPCs) now allow MD simulations based on DFT. DFTMD is sometime used interchangeably with ab initio MD. However, we consider DFTMD is a more suitable name which specifies clearly the underlying electronic structure method. The propagation of atoms in DFTMD is often treated with a classical formalism such as Newtonian or Lagrangian mechanics, but for each such time frame the equations of density functional theory (for the chosen functional) is solved for the present position of the atomic coordinates. 

The idea to combine electronic structure calculation and MD simulation seems straightforward and simple. However,  simulations using the last optimized wavefunction as the initial guess for the next time step showed large drifts in total energy even for very tight self-consistent field (SCF) convergence criteria~\cite{Hutter:2011gl}. It was realized latter that the incomplete SCF convergence leads to cumulative errors in the energy and forces which deviates from the Born-Oppenheimer (BO) potential surface in the long run. Therefore, what Car and Parrinello introduced in their seminal paper of DFTMD is the coupled equations of nuclear and electronic degrees of freedom, which ensures the time reversibility of both electron and nuclei dynamics~\cite{Car:1985ix}. This also triggered the development of the second generation Car-Parinello MD (CPMD) which attempts to combine the large time step of BOMD and the efficiency of CPMD~\cite{Kuhne:2007df}. 

DFTMD can produce physical observables beyond what classical force field-based MD (see the Section 2.2) can offer. Examples include electrochemical systems where the quantum properties of electrons play a large role in determining the total energy.  Within the context of electrochemical energy storage, DFTMD has been applied to study ion diffusion properties \cite{Perez2022}, ion coordination structures, surface pKa values and redox potentials. In addition, DFTMD can also be used for the determination of reaction free energies, as these can not always be well approximated by harmonic corrections to static single point calculations \cite{Marzari2021}.

On the other hand, due to the demanding electronic structure calculations,  the time scale of DFTMD is limited to tens of picoseconds for routine simulations on HPCs, simulating systems with up to around 1000 atoms. Slower processes such as the formation of a double layer and the charging dynamics are not within reach. Therefore, machine learning force field (see the Section 2.3) that can provide very accuracy force predictions with only a fractional cost of the corresponding DFT calculations will be of great help to tackle these time-scale and length-scale challenges.

\subsection{Empirical force field-based molecular dynamics simulations}

Electronic structure calculations can help us determine the interaction energies and optimized geometry of individual molecules and calculate electronic properties of molecules and materials. What we lack are techniques that can be used to explore the configurational space of a chemical or materials system. In this regard, molecular simulation methods based on the principles of statistical mechanics are quite useful in studying the time evolution and dynamics of complex systems at microscopic scale and estimating physical properties at macroscopic scale. 

\begin{figure}
    \centering
    \includegraphics[width=0.9\textwidth]{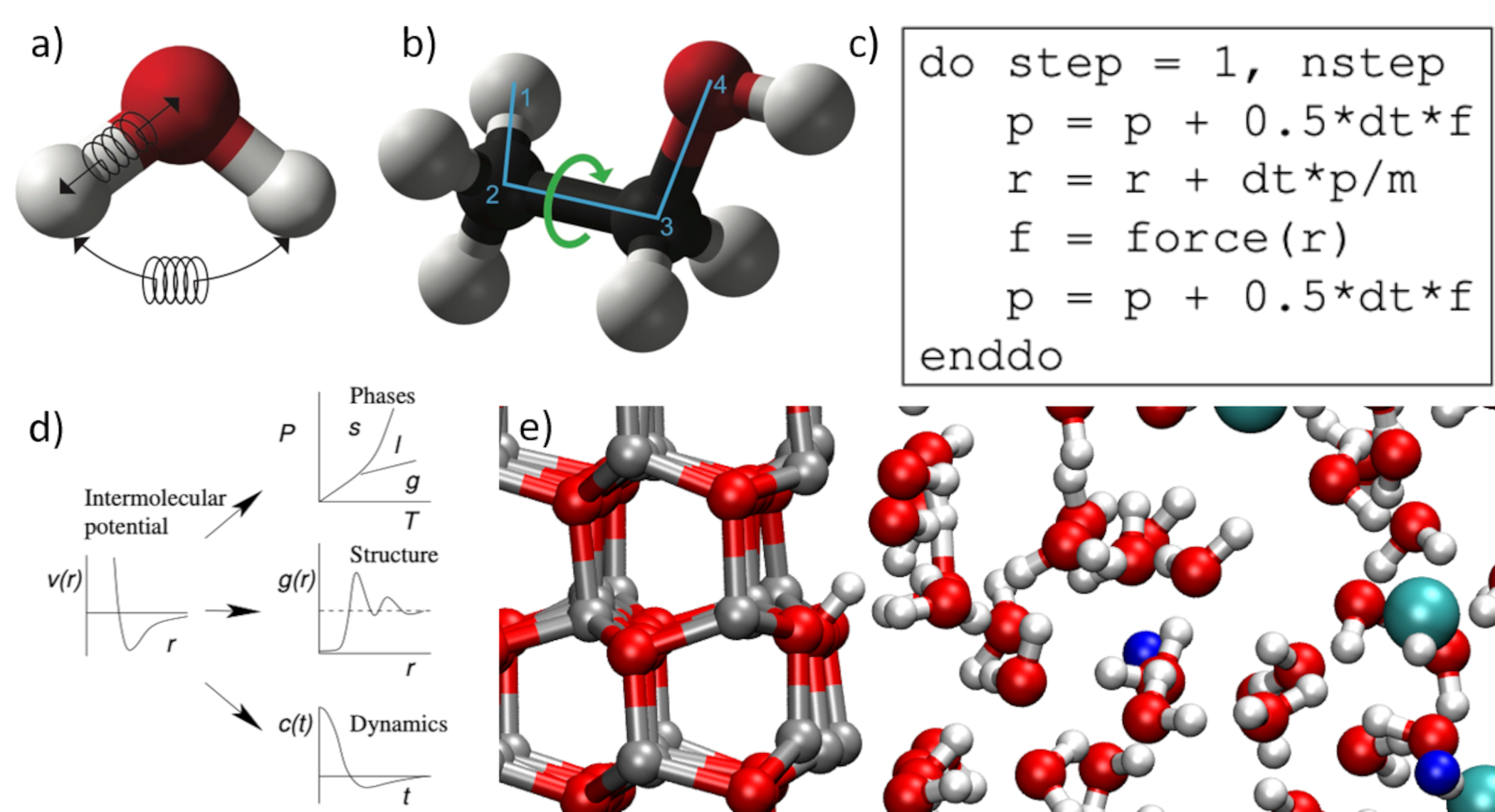}
    \caption{Empirical force-fields include (a) harmonic bond stretching and angle bending and (b) torsional rotation around dihedral angles (indicated by green arrow). Adapted from the Ref. \cite{braun_best_2019} under the CC-BY licence (Copyright 2018 Braun et al.) (c) Simple MD algorithm illustrated using a pseudo-code. Adapted from the Ref. \cite{Allen_computational_2004} (d) Insights into simulation acting as a bridge between microscopic and macroscopic properties. Adapted from the Ref. \cite{Allen_computational_2004} (e) DFTMD simulation snapshot of the \ce{ZnO}(10$\Bar{1}$0)/aq. \ce{NaCl} solution interface.}
    \label{fig:MD}
\end{figure}

Here, We will focus on MD\cite{alder_phase_1957} simulation considers time evolution of the system of atoms/molecules where every step or ‘snapshots’ along the motion of atoms/molecules the property of interest is computed and then averaged over all the time steps to obtain time average.  Time averaging and ensemble averaging are equivalent if a dynamical system is ergodic.~\cite{alavi_statistical_2011} A particle-based description of the system is considered where the initial positions, velocities and interaction energies of all the atoms in the system need to be defined. The force acting on each atom due to its interactions with other atoms can be calculated by differentiating the interaction energy present in the whole system (Figs.~\ref{fig:MD}a and ~\ref{fig:MD}b). Assuming nuclei as a classical particle, the force on each atom can be used to compute its acceleration using Newton’s second law of motion. Integration of the equations of motion with time-reversible integrator will lead to an updated  position, and subsequent time steps leading to time evolution of the system (Fig.~\ref{fig:MD}c).\cite{allen_computer_2017}. This allows us to access structural, dynamic, and thermodynamic properties of the system simultaneously (Fig.~\ref{fig:MD}d).

One of the most important aspects in MD simulations is to define the interaction energy of all the particles in the system. In a quantum mechanics description (QM), the electronic structure is treated explicitly by solving the time-independent Schr\"odinger equation, as in DFT and DFTMD methods mentioned before (also see Fig.~\ref{fig:MD}e). Despite higher accuracy, the major bottleneck is the computational cost, and as a result, limited size and time scales can be achieved. In a molecular mechanics (MM) description, molecules are represented as atoms or group of atoms within the Born-Oppenheimer approximation and the potential energy function can be written as a function of nuclear coordinates only without including the electronic degree of freedom explicitly. Each atom is assigned a charge, and the potential energy functions are parameterized by fitting to either experiment or quantum chemical calculation to calculate the bonded and non-bonded interactions usually termed as ‘force-field’. Employing periodic boundary conditions (PBC), this allows rapid simulations of 100000s of atoms at nanosecond timescale of condensed phase systems involving both bulk phase and interface. Choice of the force-field or interaction model, along with the use of constraints, treatment of cut-offs, and other simulation settings (e.g. thermostat and barostat) have to be carefully considered, which otherwise can lead to artifacts. 

In molecular force-field, bonded interactions involve bond stretching, angle bending, and torsional angle rotation. Non-bonded interactions crucially encompass all electrostatic and van der Waals interactions between atoms, whether they belong to different molecules or are within the same molecule but separated by at least three bonds. An example for the functional form in a non-polarizable force-field can be seen in equation (1) where $E(\textbf{r}^N)$ denotes the potential energy which is function of the positions of $N$ particles. 
\begin{multline}
    E(\textbf{r}^N) = \sum_{\mathrm{bonds}} \frac{k_r}{2} (r_{ij} - r_0)^2 + \sum_{\mathrm{angles}} \frac{k_\theta}{2} (\theta_{ijk} - \theta_{0})^2 + \sum_{\mathrm{torsions}} \frac{V_n}{2} (1+\cos(n\omega_{ijkl}-\gamma)) \\
    + \sum_{i=1}^N \sum_{j=i+1}^N \left( 4\epsilon_{ij} \Bigg[ \left(\frac{\sigma_{ij}}{r_{ij}} \right)^{12} - \left(\frac{\sigma_{ij}}{r_{ij}}\right)^{6} \Bigg] + \frac{q_iq_j}{4\pi \epsilon_0 r_{ij}}\right)
\end{multline}

The first term is a sum over all the bonds, with one term for every pair $ij$ of directly connected atoms and equilibrium bond length $r_0$. For each set of three connected atoms $ijk$, second term represents the sum over all bond angles. Thus bond lengths and bond angles are approximated with harmonic potentials (with force constants $k_r$ and $k_\theta$) close to the equilibrium values. The third term involves four connected atoms $ijkl$ and is the sum over all torsions with cosine functions considered owing to their periodicity. This term can also involve improper torsions where all the atoms may not be directly connected by covalent bonds. (Figure \ref{fig:MD}) The fourth term includes long-range electrostatic interactions and the short-range Lennard-Jones potential as determined by the parameters $\sigma$ and $\epsilon$. Additional or different functional forms may be included for each of these terms depending on the complexity or requirement of the system. 

There has been steady progress in the developments of force-field which started from just parameterization for simple organic molecules, but today can deal with much more complex systems including inorganic materials, metals and solid-liquid interfaces. Several packages available today to perform MD utilizing various force-fields are CHARMM \cite{brooks_charmm_2009}, AMBER \cite{case_ambertools_2023}, GROMOS \cite{scott_gromos_1999} etc., and various visualization tools such as VMD \cite{humphrey_vmd_1996}, Travis \cite{brehm_travisfree_2020}, etc are also available. Nevertheless, most of these force field can only describe fixed topology of molecules and materials and no chemical reactions are allowed with exceptions such as ReaxFF\cite{van_duin_reaxff_2001}. Interested readers can refer to Refs. \cite{braun_best_2019, gonzalez_force_2011} for more discussions on the best practices of MD simulations.

Statistical analysis of trajectories encountered during such MD simulations can be analyzed in terms of thermal quantities, such as free energy. Free energy differences allows us to determine many meaningful quantities related to phase or chemical equilibria, such as the location of a phase transition, the relative binding affinities of ligands to protein receptors, relative solvation free energy, relative p$K_\mathrm{a}$ values, concentration of point defects in crystalline solids, and so on. Unfortunately, these thermal quantities are related to the volume of the accessible phase space available rather than to an average over this space. Since MD or MC simulations usually adequately sample low energy regions of phase space, special techniques have been devised to include important high energy regions of phase space or capture "rare events" during the course of the simulation. The standard methods for determination of free-energy differences is based on free energy perturbation (FEP) method, methods based on probability distributions and histograms, and thermodynamic integration (TI). Comprehensive theory and applications behind these methods can be found in these references. \cite{chipot_free_2007}

\subsection{Machine-learning force field and generative AI}
In recent years, as the emergence of massive data sets in chemistry and material science, tremendous efforts have been made to build ML models for physico-chemical property prediction, extracting structure-activity relationships and molecular discovery.\cite{keith2021combining}. Especially, predicting forces and total energies of any conformations for a given system and generating new molecules with desired properties are two most promising directions.\cite{unke2021machine,shao2021modelling} In the former case, the machine learning force field (MLFF) trained on QM-level data narrow down the gap in computational costs between DFTMD and classical force fields, which allows us to run long-time reactive MD simulations with high-quality potential energy surface (PES). In the latter case, the rapid development of deep learning provides some powerful generative artificial intelligence (AI) models to tackle the inverse-design problem.
	
\subsubsection{Machine-learning force field}
The construction of MLFF is a supervised learning task in which the inputs are atomic structures and the output are total energy and forces. As any ML problem, it requires data curation and descriptor engineering. Kernel-based Gaussian process regression (GPR) and neural networks (NN) are common ML methods for this kind of task.\cite{bishop2006pattern}

Sufficient high-quality data is a primary condition for successful training of reliable and generic ML models.\cite{pinheiro2021choosing} The data points for MLFF can be generated from molecular simulations, and each of them is composited by descriptors and reference labels of a snapshot (e.g. total potential energy, atomic forces and stresses).\cite{shao2020pinn} The plain DFTMD is a good choice to collect the initial dataset around the local basin of PES.\cite{unke2021machine} To extend the sampling regions, one can run DFTMD simulations at different phase points (e.g. higher temperatures and pressures) or combine it with enhanced sampling techniques. Even so, there is a risk for the trained MLFF entering its extrapolation regime in which unphysical predictions will be made. An effective strategy to avoid this issue is combining MLFF with active learning strategy (i.e. learn on-the-fly).\cite{shao2021modelling} Once the MLFF made unreliable predictions for some snapshots, new reference data points will be generated based on these snapshots and the MLFF will be re-trained. Although sounds straightforward, its practical performance relies on the evaluation metrics used in uncertainty predictions, the validation from the reference QM calculations and the workflow design for automation~\cite{shao2021modelling}. 

\begin{figure}
    \centering
    \includegraphics[width=0.85\textwidth]{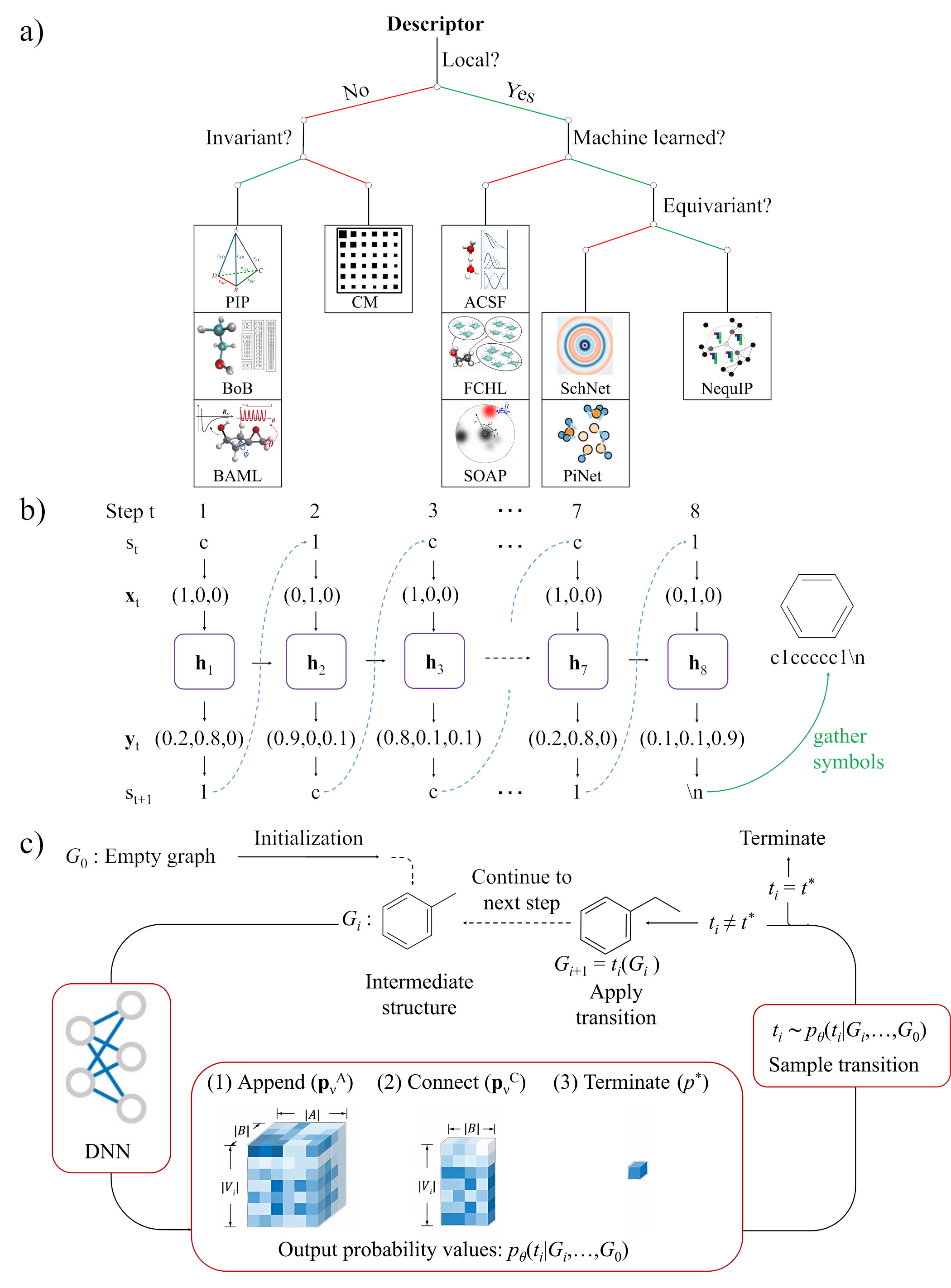}
    \caption{(a) Descriptors used in MLFFs adapted from the Ref. \cite{shao2021modelling} under the CC-BY licence (Copyright 2021 Shao et al.). In addition, Ref. \cite{huang2016communication} with permission (Copyright 2016 AIP Publishing) for BAML, Ref. \cite{batzner20223} under the CC-BY licence (Copyright 2022 Batzner et al.) for NequIP. (b) RNN-based generative AI models adapted from Ref. \cite{segler2018generating} with permission (Copyright 2017 American Chemical Society). (c) GNN-based generative AI models adapted from Ref. \cite{li2018multi} under the CC-BY licence (Copyright 2018 Li et al.)}
    \label{fig:MLFF-GenAI}
\end{figure}

Another key aspect of MLFF is designing descriptors for local chemical environments.\cite{unke2021machine} A set of good descriptors should be one-to-one mapping from the configuration to the output total potential energies, which guarantees a higher performing MLFF if more data are supplied.\cite{shao2021modelling}. These descriptors should also comply with the physical constraints of PES, e.g. translational, rotational and permutational invariance.\cite{pinheiro2021choosing} Here we will focus on the local descriptors and in particular graph-based descriptors, which have gained lots of momentum recently and are summarized in Fig.~\ref{fig:MLFF-GenAI}a.

The idea behind local descriptors is to depict atom-centered local chemical environment within a predefined cutoff sphere.\cite{pinheiro2021choosing} According to the many-body expansion ansatz~\cite{drautz2019atomic}, the total potential energy for a given \textit{N}-atom system can be expressed as $E=\sum _i^N E_i$, in which 
\begin{equation}
	E_{i}=V^{(0)}_{i}+V^{(1)}\left(\mathbf{r}_{i}\right)+\frac{1}{2} \sum_{j} V^{(2)}\left(\mathbf{r}_{i}, \mathbf{r}_{j}\right)+\frac{1}{6} \sum_{j k} V^{(3)}\left(\mathbf{r}_{i}, \mathbf{r}_{j}, \mathbf{r}_{k}\right)+\dots 
\end{equation}

The MLFF trained by local descriptors manage to estimate the interatomic interactions for a given center atom. In this regards, Behler and Parrinello pioneered on this topic and developed the atom-centered symmetry functions (ACSFs) in which two-body and three-body interactions are described by radial and angular distributions of neighbour atoms, respectively. \cite{behler2015constructing}. Instead of using bonds and angles, an alternative way to design local descriptors is based on the atomic density. An example in this category is the smooth overlap of atomic positions (SOAP) descriptor, which was created by expanding the atomic density into a set of products of orthonormal radial basis and spherical harmonic functions (SHFs) and representing the structural difference with the power spectrum coefficients.\cite{bartok2013representing}. The third category is the graph-based descriptors \cite{shao2021modelling,unke2021machine} For a given \emph{N}-atom system, one can viewed it as a graph $G=(\left\{\mathbf{v}_{i}\right\}, \left\{\mathbf{d}_{ij}\right\}, \mathbf{u})$ whose nodes and edges are atoms and interatomic connections (e.g. bonds), where $\mathbf{v}_{i}$ is the descriptors of atom $i$, $\mathbf{d}_{ij}$ is the descriptors of edge connected atoms $i$ and $j$, $\mathbf{u}$ is the global descriptors of whole graph.\cite{chen2019graph} This graph can be easily embedded into graph convolution neural networks (GCNN), then all descriptors will be iteratively and automatically learned through the convolution layers in GCNN.\cite{shao2021modelling} An example in this type of descriptor is the SchNet architecture~\cite{schutt2018schnet}. It is worth noting that the feature vector in GCNN is usually invariant with respect to the rotation of the system. This may be sufficient for predicting the total energy but not for forces which are equivariant properties. As shown quite recently by the NequIP architecture~\cite{batzner20223}, including equivariant features can lead significant performance boost in GCNN.  

\subsubsection{Generative AI}

In the point view of statistics, distribution $p\left ( \mathcal{X} \right ) $ is needed for exploring the input space $\mathcal{X}$, and the joint probability distribution $p\left ( \mathcal{X}, \mathcal{Y}\right )$ is for searching samples with desired property $\mathcal{Y}$ .\cite{jorgensen2018deep} However, in principal, both of these two distributions are unknown. To address this problem, generative AI models are designed to learn an approximation for one of these distributions. For molecular discovery, lots of generative AI models have been developed based on the variational autoencoder (VAE), generative adversarial networks (GAN), recurrent neural networks(RNN), and graph neural networks (GNN) architectures. Here we will focus on RNN and GNN architectures which are inspired by the language models and used to build auto-regressive models for molecules and materials. The typical workflows of RNN- and GNN-based generative AI models are shown in Figs.~\ref{fig:MLFF-GenAI}b and \ref{fig:MLFF-GenAI}c.

The RNN is a powerful generative AI model for sequential data and has may successful applications for the text and music generation.\cite{sutskever2011generating} Unlike conventional feed-forward NN, the hidden layer in RNN has a memory unit which enable them to remember the past information and connect their outputs to the subsequent inputs. The typical update function of the hidden state $\mathcal{H}$ can be expressed as $\mathcal{H}_{t}=f\left(\mathcal{X}_{t},\mathcal{H}_{t-1}\right)$, where $t$ is the current time step during training.\cite{sutskever2011generating}  Due to the fact that an organic molecule can be conveniently represented by its canonicalized SMILES which is a sequential string, many RNN generative models have been successfully developed for organic molecules.\cite{blaschke2020reinvent,segler2018generating}. Nevertheless, SMILES-based RNN models are limited because of their unsatisfactory exploration ability in the chemical space and low validity of generated SMILES.\cite{mercado2021graph}

Graphs as introduced previously are more powerful representations of molecules than SMILES strings. Therefore, GNN-based generative AI models have received extensive attentions.\cite{mercado2021graph} The biggest challenge of these models is how to appropriately define the generation probabilities of molecular graphs. To achieve it, a graph usually decoded into a set of binary tuples (i.e. decoding route) $\left \{ (G_t,a_t)_{t=0,1,...,n} \right \} $, where $G_t$ is an intermediate sub-graph at step $t$, $a_t$ is an action applied on $G_{t}$ to generate $G_{t+1}$.\cite{mercado2021graph} This decoding route can be collected from a sequential process driven by graph traversal algorithms (e.g. breadth-first algorithm). Based on the reverse sequence of actions, we can generate the original molecule from an empty graph. For the convenience of implementation, the actions are simply classified into three different types: add a new specific atom, add a new specific bond between two atoms, and terminate the generation process.\cite{mercado2021graph} Likelihoods of the action trajectories can be estimated during the training stage and further be used to generate new molecules by the GNN model.\cite{li2018multi}. When they are coupled with  transfer learning (TL) and reinforcement learning (RL), property-directed molecular generation can be achieved over the large chemical space\cite{segler2018generating,li2018multi}.

\section{Applications to aqueous battery systems}

\subsection{Water-in-salt electrolytes}
The first article demonstrating the promising characteristics of water-in-salt electrolytes (WiSE) under that name was published in 2015 by Suo and co-workers \cite{Suo2015WiSE}. Cyclic voltammetry performed with stainless steel electrodes in a 21 m (molal, mol/kg) LiTFSI electrolyte revealed an electrochemical stability window of around 3.0 V (Figure \ref{fig_WiSE}a). To test the practical functionality of the electrolyte, a battery cell with \ce{LiMn2O4} cathode and \ce{Mo6S8} was assembled. This cell had an open circuit voltage of 2.3 V and showed negligible HER and OER during cycling. The enlarged electrochemical stability window can be understood by investigating the bulk WiSE electrolyte and associated interfaces in terms of structural, dynamical, energetic and redox properties. 

\subsubsection{Structure and dynamics of bulk WiSE}
While the local solvation structure of the WiSE components is important for their reactivity, the bulk structure of WiSE at the nanometre length scale is important due to its strong relationship to the transport properties of the electrolyte. In the following papers, structural components  of WiSE and their interplay at various length scales is examined and discussed.

Using a range of spectroscopic techniques as well as classical MD simulations, Borodin and co-workers \cite{Borodin2017WiSE_transport} investigated the coordination environments and nanostructuring of the originally studied 21 m \ce{LiTFSI} WiSE. One interesting finding from MD simulations was that the number of solvent separated ion pairs (SSIPs) that \ce{Li+} is participating in that at high concentrations is higher than the corresponding number for \ce{TFSI-}, meaning that \ce{Li+} on average spends more time in water-rich surroundings. For validation, the SSIP-participating fraction of \ce{TFSI-}, which is experimentally obtainable from IR spectroscopy, was compared to the fraction obtained from simulations with good agreement. As in the original study by Suo et al.~\cite{Suo2015WiSE}, the average inner solvation shell of \ce{Li+} was found to consist of two oxygen atoms from water and two from \ce{TFSI-}. However, these average values reflected a mixture of \ce{Li+} coordinated by four water molecules and \ce{Li+} coordinated by four TFSI ions. These latter \ce{Li+} and \ce{TFSI-} ions were part of water-poor ion clusters or aggregates. On the nanoscopic level, the regions of water and solvated \ce{Li+} stretched around and between a scaffold of ion-rich clusters. It was further found from MD simulations that \ce{Li+} coordinated to only water diffused three times faster than Li coordinated to only TFSI, indicating that most \ce{Li+} transport occurs through the water-rich domains. For the hydrated Li ions, the vehicular mechanism appeared to dominate over the solvent exchange mechanism, as the average distance travelled by \ce{Li+} during its association with a certain hydration shell was around 5-6 Å. 

Lim et al. \cite{Lim2018water_channels} performed 2D-IR spectroscopy on and MD simulations of highly concentrated LiTFSI. The MD picture agreed with that of Borodin et al. in that a network of water-rich channels is surrounding a porous scaffold of ion clusters/aggregates and that this structuring is on the order of nanometres (see a representative structure in Figure \ref{fig_WiSE}b). Regarding \ce{Li+} transport, the simulations captured \ce{Li+} ions moving easily through the channels, with visible displacement between sequential picosecond snapshots. On the other hand, most \ce{Li} ions in the ion clusters appeared stationary during the same time period. A follow-up calculation of diffusion coefficients revealed this figure to be eight times higher for \ce{Li+} ions in the water channels compared to those forming ion clusters (0.385 vs 0.05 Åm$^2$/s). Yu et al. \cite{Yu2020_aggregate_composition} focused on the structure of these ion clusters in their 2020 computational study. As one of the validation tests of the force field used, the researchers compared the structure factor obtained by MD simulations with the one obtained by small-angle X-ray scattering (SAXS). The structure factors agreed well, showing a so-called charge ordering peak at around 10 nm$^{-1}$, and a second, shouldering peak at around 15 nm$^{-1}$. This second peak was found to be dominated by interactions between F-C and F-F atoms in TFSI, and the close correlation distance (about 0.42 nm) is evidence for aggregates with a high density of these anions. From MD simulations the group confirmed the organisation of water into small domains for a 20 m \ce{LiTFSI} solution. 

Focusing on ion transport, this study of Li and co-workers \cite{Li2019_WiSE_transport} included a comparison of four different force fields. By comparing self-diffusion coefficients, conductivities, and viscosities to experimental measurements, the authors decided on using the parameters for ionic species taken from ref. \cite{Lopes2004_FF} with the charges scaled by 0.8. As part of their results, the researchers found that the motion of cations and anions were anti-correlated for all examined concentrations, as were the \ce{TFSI} anions and water molecules. Between \ce{Li+} and water molecules, the correlation was positive, agreeing with the generally known strong binding between water and small cations. Taken together, the study found that \ce{LiTFSI} WiSE transport in terms of transport is similar to ionic liquids. On the other hand, the \ce{Li+} transference number was found to be larger than in these other electrolytes with similarly high ionic content, which is thought to be due to water-rich channels facilitating fast \ce{Li+} transport around the \ce{TFSI-} rich domains. 

A different picture of bulk WiSE structure was given by Zhang and co-workers \cite{Zhang2021_LiTFSI_structure}. In their study, classical MD was used to investigate the structure of a highly concentrated LiTSFI electrolyte. According to the authors, agreement between simulation and experiment in terms of density, structure factor, and solvation environments was so-far unmatched.
This agreement provided merit to the atomic structure produced by the simulations. Average coordination numbers were similar to those of previously mentioned studies in that \ce{Li+} on average was coordinated to two water molecules and two oxygen atoms from \ce{TSFI-} ions. Unlike these previous studies, however, \ce{Li+}-\ce{H2O} and \ce{Li+}-\ce{TFSI-} coordination number distributions were not partitioned into two populations but were both centred around two. The proposed solvent structure is thus much more homogeneous than what is suggested in previous paragraphs, and can be described as a connected TFSI network with small water clusters and \ce{Li}-ions evenly distributed throughout.

As a continuation of the above presented article, Zhang and co-workers \cite{Zhang2021_LiTFSI_dynamics} studied the transport properties of the system, in particular with focus on \ce{Li+}. Again, the force field parameters employed replicated experimental values well, this time with respect to viscosity, self-diffusion coefficients, and ionic conductivities. Experimentally, the transference number of \ce{Li+} at higher electrolyte concentration is generally found to increase. From the results of this study, the increase is attributed to the drastic immobilisation of \ce{TFSI-}  at high ionic concentrations. Furthermore, the transport mechanism of Li+ was studied. Comparing residence times for association with \ce{TFSI-} with self-diffusion coefficients, it was concluded that Li+ transfer occur primarily with the hopping mechanism.

\subsubsection{Interfacial structure of WiSE}

One of the contributing factors to the electrochemical stability of WiSE is the simple effect of lower water concentrations at the interfaces. To confirm this scarceness of interfacial water molecules, and also to obtain broader knowledge of WiSE in general, several authors have been performing computational studies of WiSE-electrode interfaces.

Vatamanu and Borodin \cite{Vatamanu2017_WiSE} provided a MD investigation about the interfacial structure of water-in-salt electrolytes. Their model system was 21 mol/kg  \ce{LiTFSI} + 7 mol/kg \ce{LiOTF} between two polarisable graphite electrode surfaces. MD simulations were run at different potentials, centred around the potential of zero charge (PZC, -0.57V vs SHE). The researchers found clear structuring at the interface, visible even a couple of nanometres towards the bulk solution. While the authors discussed and examined the properties of all species at positive, zero, and negative potentials vs the PZC, a few points were especially interesting for the electrochemical stability of WiSE. It was found that electrodes with a positive potential of 2 V vs PZC had very few water molecules in contact with the surface, as the surface was shielded by TFSI (and to a lesser degree OTF) ions. On the other hand, at a -2 V vs PZC potential, water was the dominating species at the electrode surface. 

In a study to investigate the lower limit of the lithiation potential of the anode in a WiSE, Yang and co-workers \cite{Yang2017_4V} used classical MD to simulate the behaviour of a water-in-bisalt (WiBE) electrolyte at a graphite interface held at a range of different potentials. The interfacial structure of the examined \ce{LiTFSI} and \ce{LiOTF} WiBE in contact with the graphite electrode was found to differ with the applied potential. At 2.5 V vs \ce{Li}, the dominating species in the inner Helmholtz plane were LiTFSI and LiOTF ion pairs/clusters, while water was sparse. If the applied potential was lowered to 0.5 V vs Li, water was instead the dominating species (snapshots of MD simulations at these potentials are shown in Figure \ref{fig_WiSE}c). Evidently, the negatively charged surface is dispelling anions in favour of the neutral water molecules. Thus, at the \ce{Li+} insertion potential of graphene, well below the experimental hydrogen evolution limit of 1.7-1.9 V vs Li for WiSE electrolyte, hydrogen reduction is no longer hindered by a low concentration of water in the IHP. 

Another aspect of WiSE that has been studied computationally is the double layer and its associated differential capacitance. Dhattarwal and Kashyap \cite{Dhattarwal2023_interfacial_structure} used a two-electrode setup under constant potential classical MD simulations to measure this quantity, as well as to see how this and interfacial populations changed with the addition of the ionic liquid EmimTFSI. For the 20 m \ce{LiTFSI} solution, the differential capacitance was higher (7.50 $\mu$F/cm$^2$) for the negative electrode than for the positive (4.14 $\mu$F/cm$^2$). Addition of the ionic liquid eliminated this difference. It is also found that addition of the ionic liquid reduced the concentration of water at both electrodes, which is thought to be favourable for the electrochemical stability window. In the pure WiSE the interfacial water was predominantly in the solvation shell of \ce{Li+}. Working in the same direction, McEldrew et al. \cite{McEldrew2018_continuum} published a continuum model of the WiSE double layer structure, where parameters were obtained mainly from MD simulations. The model could reproduce the qualitative and to some degree quantitative accumulation and depletion of the components of a WiSE at an interface across a range of surface charge densities. 

 The role of interfacial water in WiSE systems was investigated by Bouchal and co-workers \cite{Bouchal2020_precipitation}. Linear polarisation voltammetry produced two reduction peaks for a \ce{LiTFSI} WiSE. These peaks were assigned to bound (containing ions in the first coordination shell) and free (solvated only other water molecules) water, an assignment supported by relative populations given by classical MD simulations. The MD simulations were run on an electrolyte system between two graphite electrodes with a 3 V potential applied between them. At 20 m, almost exclusively water coordinating to \ce{Li+} was present at the interface. Online electrochemical mass spectrometry revealed that hydrogen gas was produced at a ratio to the current that suggest HER is almost solely responsible for the first reduction peak. Thus, at a potential where the SEI was observed to form without the reduction of TFSI. These measurements and simulations pointed out that the free water that gets oxidized in WiSE increases the local concentration of \ce{LiTFSI}, causing it to precipitate. The formed layer was too viscous to allow more water molecules to reach the electrode. In addition, the authors found that water reduction product cause chemical degradation of TFSI ions. Since this process was also observed in solutions of lower concentration than 21 m, the precipitation is required for the SEI formation.

\subsubsection{Redox potential and acidity of WiSE}

Besides the reduced concentrations of interfacial water, other potential mechanisms behind the widening of the electrochemical stability window have also been studied computationally. These mechanisms are often based on changes in the inherent thermodynamic reduction potential of the constituent species.

In the original 2015 WiSE article\cite{Suo2015WiSE}, Suo et al. investigated the reduction potential of \ce{TFSI-} when complexed with \ce{Li+}. They performed DFT calculations to investigate possible shifts in reduction potentials. For these calculations, the authors used the long-range corrected LC-$\omega$PBE functional. A few example clusters of \ce{TFSI} and \ce{Li} ions were studied with respect to reduction potential. The solvation environment of these clusters was not explicitly represented in the calculations but modelled by the SMD continuum solvation model. Compared to isolated \ce{TFSI} with a reduction potential of 1.4 V vs \ce{Li+}/\ce{Li}, the reduction potentials of the studied clusters showed reduction potentials of up to 2.9 V vs \ce{Li}. These values lie close to, and even above the reduction potential of neutral water (2.63 V vs \ce{Li+}/\ce{Li}). According to these calculations, a sufficient reduction of TFSI anions to form a passivating SEI is possible before large currents due to hydrogen evolution develop. This conclusion was further supported by DFT calculations of the density of states (DOS) of 5 and 22 m \ce{LiTFSI} solutions at the level of theory of hybrid functional HSE06.

\begin{figure}
    \centering
    \includegraphics[scale=0.9]{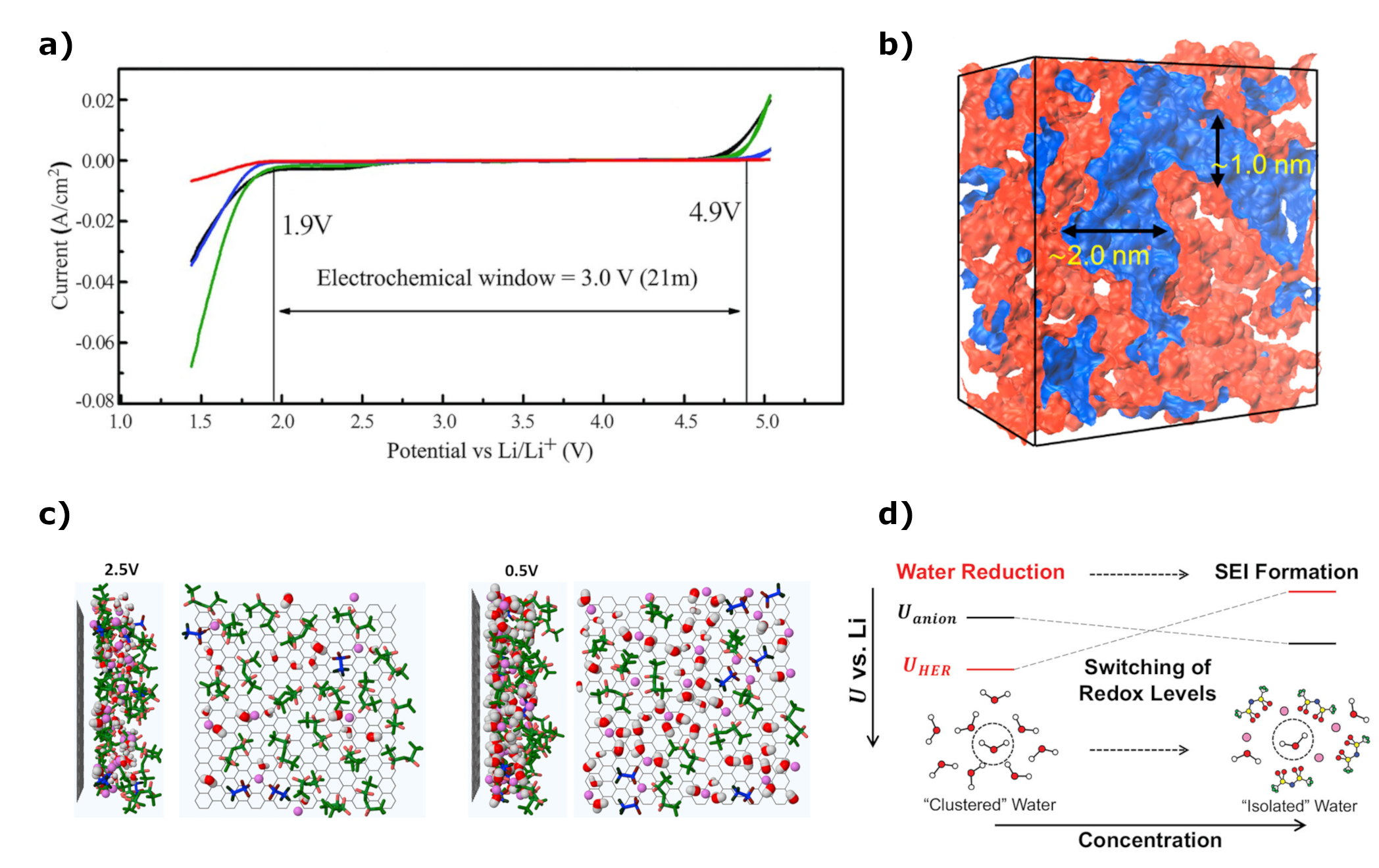}
    \caption{ (a) The impressively high electrochemical stability window of 21 m \ce{LiTFSI}. Adapted with permission from   \cite{Suo2015WiSE} (Copyright 2015 American Association for the Advancement of Science). (b) Simulated structure of bulk LiTFSI WiSE (21 m). This simulation suggests nanomeric water channels intertwined with ion cluster networks. Reprinted with permission from  \cite{Lim2018water_channels} (Copyright 2018 American Chemical Society). (c) Water being largely excluded from the (graphite) electrode surface in a \ce{LiTFSI}+\ce{LiOTF} water-in-bisalt electrolyte at an applied potential of 2.5 V vs Li., inhibiting hydrogen evolution. Reprinted with permission from  \cite{Yang2017_4V} (Copyright 2017 Elsevier). (d) The effect of ion-dominated coordination spheres on redox levels in a concentrated LiTFSI solution. Reprinted with permission from  \cite{Wang2022_switching} (Copyright 2022 American Chemical Society).}
    \label{fig_WiSE}
\end{figure}

Wang \cite{Wang2022_switching} and co-workers instead used a combination of machine-learning MD (MLMD) and DFTMD for an examination of the reduction potentials in question. The researchers were building on earlier studies, by themselves and others, showing favourable shifts in orbital energies to shift water reduction to higher potentials, and wanted to deepen the study by calculating the full free energies of water- and \ce{TFSI-} reduction. Their own density of states-studies were based on MLMD calculations, which allow the group to obtain a nanosecond long trajectory with an accuracy that ought to exceed that of classical MD simulations and also allow the collection of electronic structure information along the trajectory. Obtained DOS plots showed the LUMO of the electrolyte solution located on \ce{TFSI-} already at concentrations of 5.24 m. Moving on to the calculation of free energies to obtain thermodynamic reduction potentials, the method was switched to DFTMD with BLYP functional, with energy statistics for the reduction potential calculated by the more accurate hybrid functional HSE06. In order to obtain well referenced reduction potentials, the so-called computational \ce{Li} reference electrode was used, following a methodology first applied to the standard hydrogen electrode by Cheng and Sprik \cite{Cheng2012_SHE}. Using this method, it was found that the reduction potential of \ce{TFSI-} did not change much over a range of concentrations from 0.64 to 25.6 m. On the other hand, the potential of hydrogen evolution vs the Li reference electrode was found to decrease with around 0.7 V over the concentration range (see the scheme in Figure \ref{fig_WiSE}d). 
 
Proton activity also plays an important role in the electrochemical stability of WiSE. Yokoyama et al. \cite{Yokoyama2018_electrochemical_stability} in 2018 suggested that the HER and OER moderate the local pH at the interfaces in such a way that they lower their own thermodynamic driving force. Neutral water solutions at low to moderate salt (or acid/base) concentrations have larger electrochemical stability windows than acidic or basic, prompting the group to suggest that \ce{H+} or \ce{OH-} formed at the interfaces slow down the oxidation or reduction unless the concentration of the complimentary species (corresponding to basic or acidic environment) is high. Their theory was tested by adding buffer to the solution, which turned out to reduce the stability window to the HER and OER potentials given by the acidic and basic solutions, respectively. Additional chemical insights into the local pH of WiSE came from Han et al. \cite{Han2020_unusual_acidity}, who experimentally found a strong acidity (pH 2.4 for 20 m solution) of a LiTFSI WiSE. The group used DFT calculations, classical MD, and NMR spectroscopy to investigate the origin of this acidity. While p$K_a$ values of water bound to \ce{Li+} or to one \ce{Li}-\ce{TFSI} contact ion pair were found to be high (12.18 and 15.37), the two investigated clusters with water bound to two \ce{Li}-\ce{TFSI} contact ion pairs had drastically lower p\ce{K_a} values of 0.93 and 1.73. From the group's classical MD studies of the \ce{LiTFSI} WiSE molecular structure, the two-fold coordination to \ce{Li+}-\ce{TFSI-} was found to have a high probability of occurring, though it was of course not possible to fully compare the whole solution to the two selected clusters.  

\subsection{Layered transition metal oxides}
Many cathodes, or positive electrodes, that has been investigated for use in aqueous batteries have been composed of transition metal oxides. Charge storage mechanisms in these compounds are based on cation (including proton) intercalation, together with reduction of the oxide. In the search for optimal metal oxide cathodes, many computational studies have been undertaken.
The overview of computational studies and findings here will focus on layered metal oxides, but other polymorphs will also be mentioned when necessary. 

\subsubsection{Cobalt oxides}

As with many oxides, the use of \ce{CoO2} in aqueous batteries is hindered by its dissolution in water. Posada-Pérez, Hautier, and Rignanese \cite{Posada2022_proton_LiCoO2} used DFT calculations to examine how the adsorption of protons lead to oxygen vacancy formation and thus to the dissolution of the material. Their study was DFT+U based and used the PBE functional to explore the interaction of protons with the four lowest-energy surface terminations: (001) and (012) (which are polar), and (104) and (110) (which are nonpolar). For these surfaces, varying degrees of lithiation was modelled. Free energies of surfaces and bulk structures were approximated by internal energies calculated by DFT, for some species complemented by zero-point vibrational energies. Using the same methodology, energies of proton adsorption was calculated for different sites and with different surface coverage, before O vacancy formation. Finally, the energy of oxygen vacancy formation was calculated by removing a OH group from the surface where \ce{H+} had been adsorbed.  Based on their findings, the authors included that that oxygen vacancy formation due to proton adsorption does not play a crucial role in the deterioration of cobalt oxide layered electrodes in aqueous solutions.  

George and co-workers \cite{George2024_OER_LiCoO2} investigated water adsorption and the oxygen evolution reaction (OER) on the \ce{LiCoO2} cathode. DFT+U calculations with the PBE functional were used. The surfaces included in this study were also the (001), (012), (104) and (110) terminations. Surfaces with varying fractions of \ce{Li+} coverage were compared in order to gain insight into the energetics at the full range of cycling stages. Energies of the oxygen evolution reaction steps were related to experimental values using the computational hydrogen electrode method from Nørskov and co-workers \cite{Norskov2004}. Two mechanisms for OER were considered. The first one was denoted simply ``OER" and have the oxygen atoms of the final molecule come from two sequentially adsorbed water molecules. The second mechanism was called the lattice oxygen evolution reaction (LOER) and involves one oxygen atom being supplied by the oxide lattice. The researchers found that oxygen evolution was predicted to have very high overpotentials for all surfaces and both mechanisms, except for a few exceptions, where the lowest overpotential registered was 0.56 V for the LOER at the \ce{Li_{0.75}CoO2} (104) surface. Thus, oxygen evolution is not predicted to constitute a large problem during aqueous battery operation. However, on the other hand, the first steps of the process required little overpotential, resulting in hydroxyl formation which might form interfacial cobalt and lithium hydroxides (see Figure \ref{fig_oxide}a for an example of the calculated energy levels of the intermediates). 

Oh and co-workers \cite{Oh2023_interfacial_layers} studied the interfacial structure of layered cobalt oxides (LCO) in aqueous solutions of different \ce{Li}-salts under different charge states of the electrode. MD simulations were run, where DFT-based solute molecules were propagated in an average solvent electrostatic potential. For this study, the DFT calculations were performed with the PBE functional. From running MD simulations of different salts (0.5 M \ce{Li2SO4}, 1 M \ce{LiNO3}, 1M \ce{LiClO4}, and 1 M \ce{LiTFSI}) it was found that the kosmotropic anions remained at higher concentrations close to the interface at the potential of zero charge and in the case of \ce{SO4^{2-}} even at potentials below the PZC. Sulphate anions remained at the oxide interface when a negative charge density of -11.5 $\mu$C cm$^{-2}$ was applied, appearing to be stabilised by participation in contact ion pairs with lithium ions. Experimentally, this solution also provided the highest cycling stability of the four examined solutions. As water dissociation followed by proton intercalation reduce the capacity for \ce{Li+} intercalation, a lower interfacial water concentration is also positive for cathode capacity.

\subsubsection{Manganese oxides}

Another layered metal oxide that has attracted attention for aqueous batteries is the layered polymorph of manganese oxide, known as birnessite, which together with other \ce{MnO2} polymorphs can intercalate both protons and other cations. 

\begin{figure}
    \centering
    \includegraphics[scale=0.9]{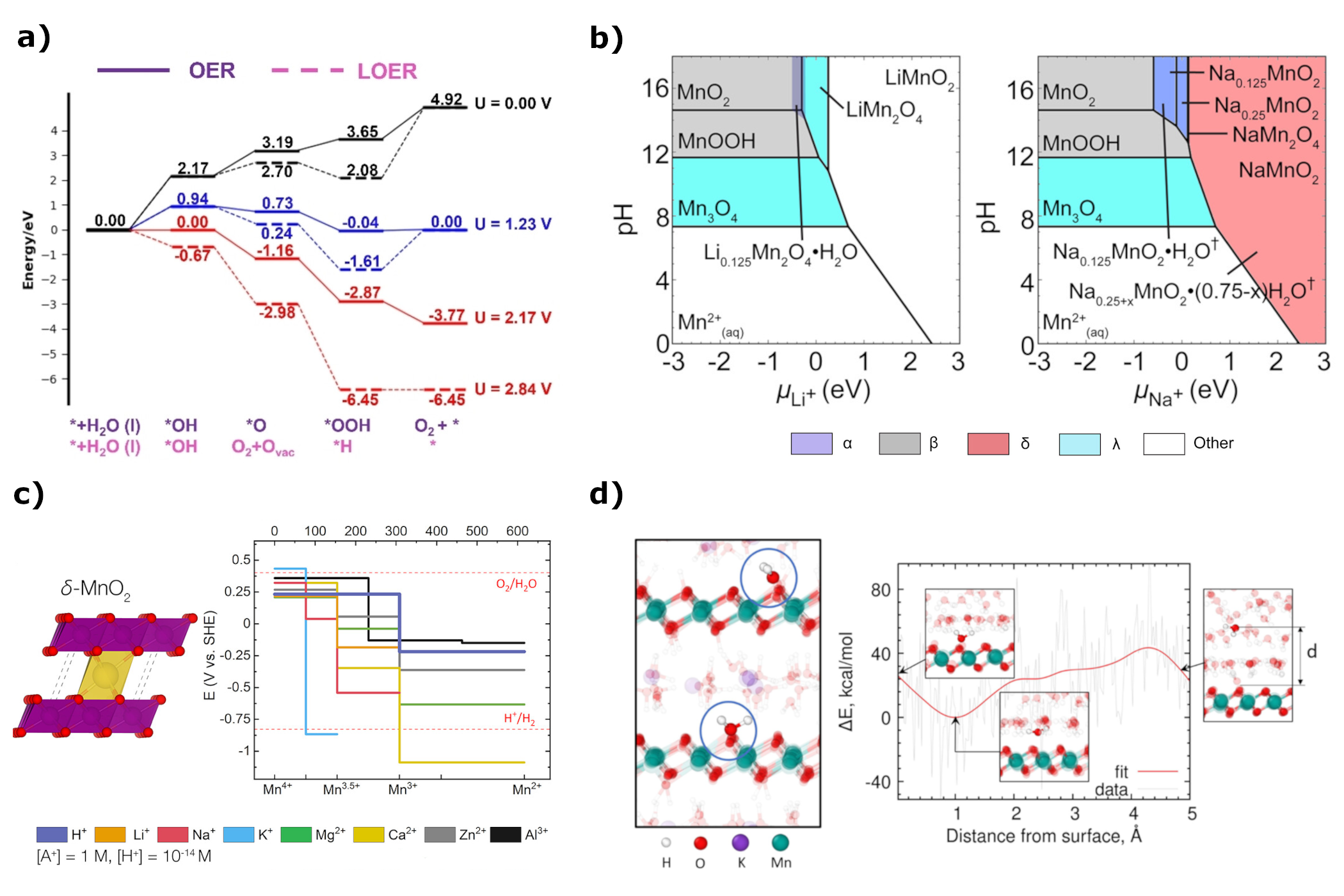}
    \caption{(a) Thermodynamic energy levels for the intermediates of oxygen evolution reaction mechanisms on \ce{Li_{0.5}CoO2} for U=0 V, 1.23V, and the least overpotential needed to make the reaction thermodynmically favourable. Reprinted from Ref. \cite{George2024_OER_LiCoO2} under the CC BY-NC-ND licence (Copyright 2024 George et al.). (b) Phase diagrams for manganese oxide-derived compounds obtained from experimental and computational data. The left phase diagram considers intercalation of \ce{Li+}, and the right intercalation of \ce{Na+}. Reprinted with permission from \cite{Kitchaev2017_MnO2_phases} (Copyright 2017 American Chemical Society). (c) Calculated insertion voltages for various cations in birnessite. Reprinted with permission from  \cite{Carlson2023_MnO2_selectivity} (Copyright 2023 American Chemical Society).(d) Energy profile of oxygen vacancy formation in birnessite. Adapted with permission from \cite{Saeed2021_birnessite_capacitance}(Copyright 2021 Wiley-VCH GmbH).}
    \label{fig_oxide}
\end{figure}

A first point of theoretical interest might be the factors determining the phase selection of \ce{MnO2}, especially as the material is known to go through phase transitions during electrochemical cycling. Kitchaev and co-workers \cite{Kitchaev2017_MnO2_phases} examined the effect of water and cation intercalation on the stability of manganese oxide polymorphs. They studied the six most common polymorphs, $\beta$, $\gamma$, R, $\alpha$, $\delta$, and $\lambda$, where the $\delta$ structure is the layered polymorph birnessite. The meta GGA-functional SCAN was used for energy computations. For cations other than protons, the conformational entropy was deemed low enough to be ignored, but for proton intercalation an entropy estimation based on an ideal lattice solution was added to the potential energy for a better estimation of the free energy.  All examined non-proton cations except \ce{Li+} were found to favour the $\delta$ polymorph at some concentration, generally in the higher region of electrochemical potential. Figure \ref{fig_oxide}b shows two phase diagrams describing the effects of \ce{Li+} and \ce{Na+} activity on the most stable phase. Without any cation intercalation, $\delta$-\ce{MnO2} was the only phase found to hydrate spontaneously. For the other phases, hydration was favourable when water molecules could interact with the intercalated cations. 

Carlson et al. \cite{Carlson2023_MnO2_selectivity} also studied intercalation into \ce{MnO2} polymorphs computationally. The first part of their study was an evaluation of DFT functionals, comparing SCAN+U and PDB+U in order to assess accuracy and speed with respect to their planned investigation. Following the investigation, the group continued their work using the PDB functional with a U value of 2.75 eV. Even though the PDB + U method did not predict the correct ($\beta$) ground state polymorph and had a larger error in lattice parameters, its errors for formation energies were slightly smaller than for the SCAN + U scheme. Free energies of intercalation were then estimated from the DFT energies by taking into account zero-point energies, phonon vibrational energies, and experimental entropy values. The authors found proton intercalation to be favoured over all other examined cations at neutral pH. At pH 14, the $\alpha$ and $\gamma$ polymorphs intercalated a couple of other ion types significantly more favourably than protons, but the other polymorphs saw cation insertion being isoenergetic or less favourable compared to proton insertion despite the low activity of the latter. (For the example of $\delta$-\ce{MnO2}, see Figure \ref{fig_oxide}c.) Regarding water co-insertion, it was found that the smaller ions (\ce{Li+}, \ce{Mg^{2+}}, \ce{Zn^{2+}}, and \ce{Al^{3+}}) were more stable with intercalated water than without in the $\alpha$ polymorph. Partial solvation shell insertion was not found to be favourable with respect to dry insertion for $\delta$ \ce{MnO2} and any of the considered cations. However, water co-insertion was still observed experimentally for the layered structure, suggesting that unaccompanied insertion of these cations may have a high kinetic barrier.

Differences between different polymorphs are also important for the potential use of manganese oxide as a cathode in proton batteries. To this end, Dai et al. \cite{Dai2023_MnO2_proton} studied the same six \ce{MnO2} polymorphs using both experimental and computational methodologies. When proton insertion energy was calculated by PBE-DFT, the lowest stabilisation was found for the $\delta$ polymorph. Bader charge analysis suggested that the \ce{Mn} ions were reduced the least at the insertion of a proton (change in Bader change -0.09 \%), which the authors believed to be the reason for the low intercalation energy. Subsequently, the volume change due to the Jahn-Teller effect on \ce{Mn^{3+}} upon proton insertion was compared. Here, the $\delta$ polymorph showed the smallest change, which ought to be positive for structural stability during cycling. Experimentally, however, the $\delta$ polymorph showed poor cycling stability, which is suggested to originate in the layered structure and absence of a 3D covalent network. 

Intercalation of cations in \ce{MnO2} is fast enough to give a capacitive response. Saeed et al. \cite{Saeed2021_birnessite_capacitance} studied this process, specifically in an attempt to find out if proton coupled electron transfer (PCET) contributes to the capacitive charge or not. Computationally, the method used was grand canonical Monte Carlo (GCMC) based on the reaxFF force field. The GCMC scheme allowed for insertion of protons in the interlayer, as well as vacancy formation upon protonation of oxygen. The authors found that proton intercalation sometimes, but not always, lead to oxygen vacancy formation. The energy profile of oxygen vacancy formation is presented in Figure \ref{fig_oxide}. In a follow-up study by the same group \cite{Boyd2021_birnessite_interlayer}, the nature of cation intercalation in birnessite was investigated using both experimental techniques, MD and MC simulations, and DFT calculations. GCMC calculations showed that reduction of the host material led to intercalation of both \ce{K+} and water. At the same time, DFT calculations showed that intercalated \ce{K+} had its energy minimum in the middle of the two oxides layers, thus not being closely coordinated to any of them. Together with the experimental results, the combined computational techniques suggest that cation intercalation primarily occurs with a partial solvation shell, thus showing an adsorption behaviour lying between specific and non-specific adsorption.

\subsubsection{Ruthenium oxides}

One material that has shown impressive proton storage qualities is \ce{RuO2}, which so far has been studied mostly with focus on its pseudocapacitive properties. One computational study of \ce{RuO2} was performed by Ozolinš, Zhou, and Asta \cite{Ozolins2013_ruthenia_protons}, where several aspects of the material was studied using ab-initio methods. For example, the researchers studied the electronic structure and the charge transfer effects upon H intercalation. They found that the commonly used description of Ru being reduced to its 3+ oxidation state is misleading, as Bader charges assigned only an extra 0.3 electrons to Ru upon hydrogen insertion. The group also found that the bulk intercalation voltage of protons was low (0.1 V) and that higher voltages (0.3-1.2 V) promoted additional surface protonation reactions. While one-dimensional diffusion was found to be easily accessible at room temperature, diffusion between different \ce{RuO2} channels had significantly higher energy barriers. Lastly, investigations into the favoured configurations of structural water suggested that regions/clusters of water were preferred to isolated water molecules. 

Parker, Robertson, and Imberti \cite{Parker2019_RuO2_surface} used elastic and inelastic neuron scattering together with DFT-studies to investigate the structure, and especially the surface, of hydrous ruthenium oxide. Their results confirmed the nanoparticulate structure of the material and the confining of water to the boundaries of the particles. By simulating inelastic neuron scattering spectra with DFT and comparing to experimental, the authors found that the surfaces of grains likely are almost fully covered with -OH groups. Additionally, there might be small amounts of water hydrogen bonded to the surface. This model suggests that proton transport in the material occur with a Grotthuss -like mechanism. 

Using computational methods, it is also possible to simulate pseudocapacitance processes and their voltage dependence. Keilbart, Okada, and Dabo \cite{Keilbart2019_RuO2_pseudocapacitance} used a multiscale approach to simulate \ce{RuO2} pseudocapacitance, taking both proton adsorption events and double layer charging into account. First, energies of a set of protonated unit cells were calculated, after which the voltage/capacitance dependence at finite temperature was studied with the help of the previously determined energies and a Monte-Carlo framework. In large part, results reproduced experimental values and trends well, although the isotherms of the (101) surface was not quantitatively comparable to experimental data. The study showed the importance of including both faradic and double layer capacitance in models, and the researchers found that subsurface adsorption of protons could increase the pseudocapacitive energy density for the (110) surface.  

\subsubsection{Tungsten oxides}

Tungsten oxide has also been examined as a proton intercalation material. Lin and co-workers \cite{Lin2014_WO2_proton_diffusion} studied the proton intercalation into and diffusion through dry and hydrated tungsten oxide. A search for low-energy intercalation sites was performed by scanning the available space inside the unit cell. The lowest energy position was found to be at a bridging oxygen between \ce{WO6}-octahedra.
Subsequently, the diffusion pathway and barriers were calculated with the nudged elastic band (NEB) method. For \ce{WO3}$\cdot$\ce{2H2O} it was found that the lowest energy diffusion path did not involve structural water but lay between diagonal bridging atoms, accompanied by slight rotation of MO6-octahedra (for a visualisation of this pathways and its energy barriers, see Figure \ref{fig_layered}b). The same pathway was found for WO3. For the monohydrate \ce{WO3}$\cdot$\ce{H2O}, both the lowest energy intercalation site and the diffusion pathway was found to be different and involve terminating oxygen atoms. Between these atoms, the proton diffusion barrier was very low (0.07) eV. However, the authors did not believe this pathway to be as favourable in a real system due to the inclusion of defects in the material and repulsive interaction between intercalated protons. To further examine this hypothesis, two protons were intercalated together in the same \ce{WO3}$\cdot$\ce{H2O} system. Examinations showed higher strain and high proton-proton repulsion. Crucially, the combined preference of the terminating oxygen sites and the increased strain was found to block both diffusion paths into the material.

A preference for proton intercalation at bridging oxygen sites is also one of the conclusions obtained from a joint computational and experimental study by Mitchell et al. \cite{Mitchell2019_WO2_hydrates}. In this study, it was also hypothesised that the favourable effect on intercalation of structural water in \ce{WO2} is mainly due to structural effects, allowing for easy structural changes in the two-dimensional layers, but stabilising the structure perpendicular to the layers.

Another tungsten compound, layered \ce{H2W2O7} was studied by Wang et al. \cite{Wang2021}. The DFT calculations performed by the group suggested that proton prefer to inhabit equatorial bridging oxygen sites at low proton content, but that the preferred site switches to the terminal oxygens. This suggested intercalation behaviour at high charge states was supported by simulated and experimentally measured XRD peaks.

\subsubsection{Other types of metal oxides}

A further group of oxides that has been studied for use in aqueous electrolytes are the vanadium oxides. Wu and co-workers \cite{Wu2019_structural_water_V2O5} performed a computational study on the effect of structural water in bilayer \ce{V2O5} with respect to \ce{Zn^{2+}} intercalation. Unsurprisingly, DFT calculations found that the interlayer spacing was increased when the structure contained interlayer water. By studying possible intercalation sites for \ce{Zn^{2+}}, the group found that interlayer water in most cases kept \ce{Zn} from interacting with both surrounding oxide layers, instead adsorbing closely to one of them. A study of the volume expansion during intercalation found that \ce{V2O5} with interlayer water saw much smaller changes in volume during \ce{Zn^{2+}} intercalation, which is favourable for material stability. Theoretical operating voltage for the cathode material was found to be 1.86 V vs Zn then I for \ce{V2O5}$\cdot$\ce{1.75H2O}. However, this value is noticeably higher than the experimentally obtained OCV of 1.3V, which the authors believe is mainly due to inaccuracies associated with the BPE functional. The average voltage across full intercalation was calculated to be 0.74 V, agreeing with the experimental voltage of 0.71 V. Electronic structure calculations suggested that interlayer water is accepting some of the negative charge during \ce{Zn^{2+}} intercalation, giving the water a deeper role than simply ``padding" the ions. When studying the diffusion pathway of \ce{Zn^{2+}} with a nudged elastic-band method, the addition of interlayer water resulted in diffusion barriers almost half as low as for the dry material. Part if this effect is thought to come from electrostatic shielding of oxide-\ce{Zn^{2+}} interaction by water, but the authors also note that the observed charge transfer between \ce{Zn^{2+}} and \ce{H2O}-\ce{O} makes the situation more complex. Finally, the theoretical capacity of \ce{Zn^{2+}} intercalation was calculated. A value of 251 mA h g$^{-1}$ was found, which is lower than the experimentally measured value of 381 mA h g$^{-1}$. According to the authors, two possible reasons for the underestimation are the added effect of double layer charging, and the possibility of \ce{Zn^{2+}} exchanging with structural water, allowing for a higher Zn loading than predicted.

\ce{MoO3} is known to go through phase changes during electrochemical cycling. Ikezawa et al. \cite{Ikezawa2023_MoO3_proton} studied the occurrence of and the mechanisms behind these phase transitions through X-ray diffraction and PBE+U DFT calculations, with the goal of using \ce{MoO3} as a positive electrode material in aqueous proton batteries. From experimental observation, it has been suggested that the first step after the first reduction and proton insertion, transition from phase I to phase III, is irreversible. Upon the first reduction from oxidized \ce{MoO3}, protons preferred to intercalate within the oxide layers rather than in the interlayer space, which is consistent with the form of phase I \ce{MnO3}. When protons were confined to the intralayers, the preferred structure showed alternating proton-filled and empty channels for \ce{H2Mo8O24}. From studies of \ce{H4Mn8O24}, i.e., corresponding to a higher proton concentration, it was found that the structure was more stable with all protons in the intralayer than in the interlayer. At the even higher concentration of protons, \ce{H8Co8O24}, the structure rearranged to show a clear preference for all the protons being in the interlayer sites. Since no geometry optimisation caused protons to switch between the intra- and interlayer positions, the diffusion boundary for this transport is believed to be high. Transition from the I phase and III phase is thus suggested to occur by phase boundary migration, and this is also the reason why other phases than I form as phase III is oxidized. 

\begin{figure}
    \centering   \includegraphics{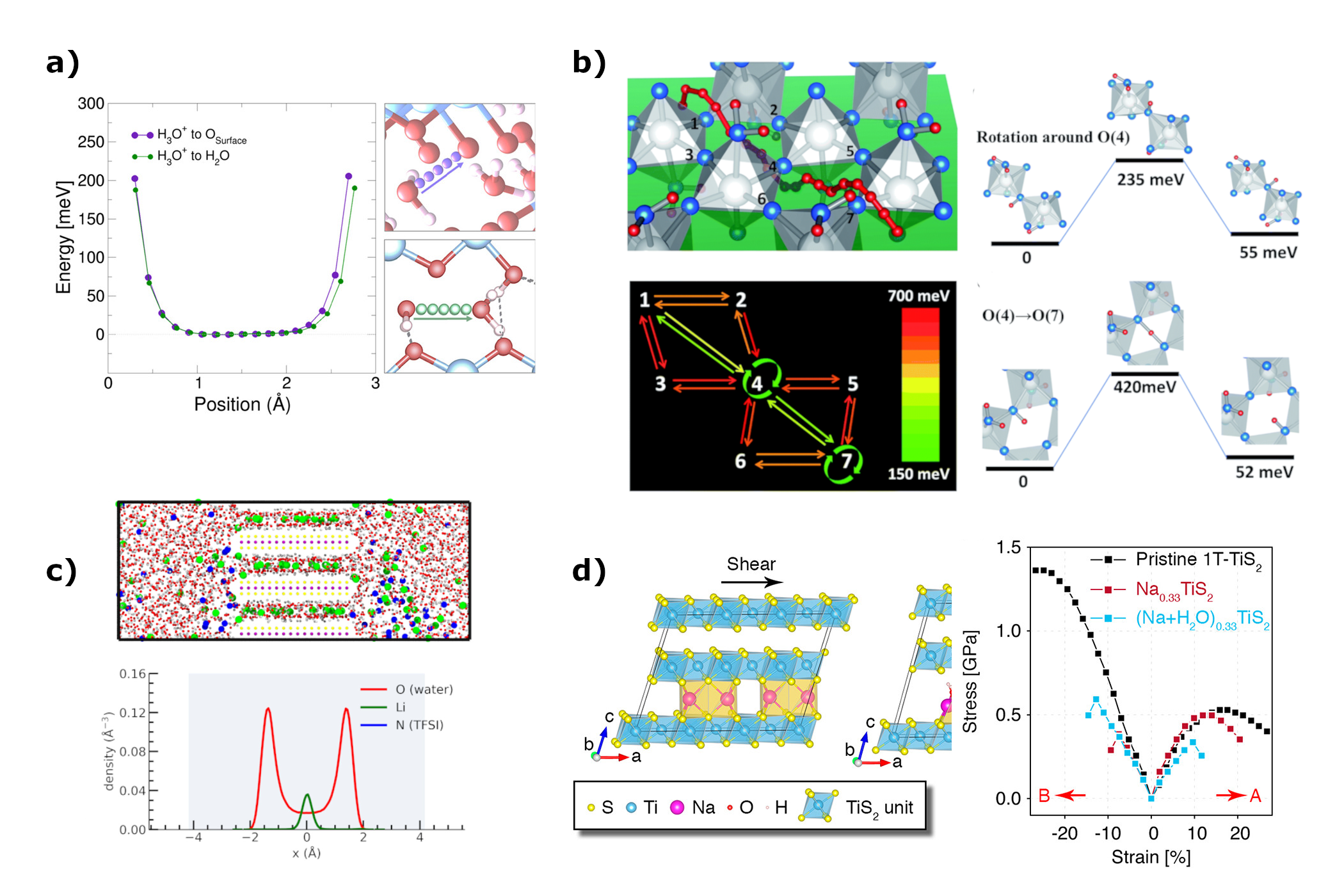}
    \caption{(a) The energy of a proton in \ce{TiO2} when passing between two intercalated water molecules. Reprinted with permission from  \cite{Kang2020_titanate_proton} (Copyright 2020 American Chemical Society). (b) Proton diffusion pathway through \ce{WO3}$\cdot$\ce{2H2O} and the encountered energy barriers. Reprinted with permission from \cite{Lin2014_WO2_proton_diffusion} (Copyright 2014 The Royal Society of Chemistry). (c) Distribution of \ce{H2O}, \ce{Li+}, and \ce{TSFI-} in the interlayeres of \ce{TiS2} at -0.65 V vs SHE. Reprinted from Ref. \cite{Zhang2024_hydration_Li_batteries} under the CC-BY licence (Copyright 2024 Zhang et al.). (d) The effect of structural water on strain in \ce{TiS2}. Reprinted with permission from  \cite{Li2020_TiS2_desalination}(Copyright 2020 American Chemical Society).}
    \label{fig_layered}
\end{figure}

An example of using DFT calculations to elucidate the properties of a newly synthesised layered structure with the approximate molecular formula (\ce{H3O})$_{1.35}$\ce{Ti3O_{6.67}}$\square_{0.33}\cdot$1.73\ce{H2O} comes from Kang \cite{Kang2020_titanate_proton} and co-workers ($\square$ denotes a vacancy). By trying a new synthesis protocol, the group synthesised a titanium oxide with tetragonal trititanate structure. The layered structure contained structural interlayer water, which is thought to promote proton diffusion. Electrochemical measurements also proved that the material had potential for proton-based charge storage. DFT calculations using the PDB functional was used to study the ionic transport through the material. Limited by computational complexity, the interlayer water and intercalated protons were modelled with two hydronium ions per supercell. By studying energy barriers for proton diffusion, it was found that the proton is relatively easily transported both between hydronium ions and between the oxygen atoms lining the walls of the interlayer space (Figure \ref{fig_layered}a).

\subsubsection{Related transition metal dichalcogenides}

There are several other categories of layered materials that might serve as electrodes in aqueous batteries and  one related class is the transition metal dichalcogenides (TMDs). As TMDs have been used as electrodes in numerous studies of Zn-ion batteries \cite{Li2022_TMD_review}, several computational works examine the materials in the role of \ce{Zn^{2+}} intercalation hosts. Researchers have examined ways of doping, hybridising, or in other ways modifying the chalcogenides in order to optimise capacity, charging rate, and structural stability. One such study was performed by Mao and co-workers \cite{Mao2023_Zn_VS2}. In this study, the effect of vanadium vacancies on \ce{Zn^{2+}} diffusion in \ce{VS2} was one of the questions targeted. The group found that vacancies both allowed for faster \ce{Zn^{2+}} transfer through the interlayers and for diffusion in the dimension perpendicular to the \ce{VS2} sheets. Li and co-workers \cite{Li2022_proton_lubricant} used DFT to evaluate the possibility of PEDOT insertion into \ce{MoS2}. After a favourable prognosis from their simulations, the group synthesised the hybrid material and continued to study it computationally with regard to \ce{Zn^{2+}} intercalation. These additional calculations showed favourable co-insertion of protons and \ce{Zn^{2+}}, as well as a shielding effect of the intercalated protons with respect to electrostatic interactions between \ce{Zn^{2+}} and the host material. The same group \cite{Li2024_gap_filling_ZnMoS2} later performed another computational study, this time to find the preferred sites of co-inserted \ce{H+} and \ce{NH4+} into a \ce{MoS2}/(reduced graphene quantum dots) material. From these simulations, co-insertion appeared beneficial for \ce{Zn^{2+}} intercalation, due to the gap-filling and electrostatic shielding effects of the additional cations.

For other cations such \ce{Li^{+}} and \ce{Na^{+}}, computational studies of TMDs aimed at aqueous battery applications have also been carried out. Recently, the common electrode material \ce{TiS2} was re-examined for use in Li-ion aqueous applications by Zhang and co-workers \cite{Zhang2024_hydration_Li_batteries}. Within this study, MD simulations were used to characterise three experimentally observed regions of the differential capacity curve. It was found that these regions corresponded to different compositions of the interlayer, mainly with respect to water co-insertion. At low potentials, two full interlayers of water were observed, providing fast transport of \ce{Li+} (the distribution of species in the interlayer can be seen in Figure \ref{fig_layered}c). In a similar vein, although with focus on different applications, Li and co-workers \cite{Li2020_TiS2_desalination} studied the intercalation of \ce{Na+} into \ce{TiS2}. As their research was directed towards water desalination, the effect of water on the intercalation process was the main point of interest. One important finding was a lower tolerance to sheer stain after sufficient co-intercalation of \ce{Na+} and \ce{H2O} (Figure \ref{fig_layered}d). This finding explained the lower maximum capacity of \ce{TiS2} in aqueous compared to dry conditions. Later, the same group used deep-neural-network-assisted MD to study water-\ce{TiS2} interactions at interfaces, with focus on proton transfer \cite{Li2023_TiS2_interfaces}. Their findings suggest that the possibility of proton transfer from water to \ce{TiS2} is highly dependent on surface morphology. 

 MXenes, which are transition metal carbides or nitrides, are also promising to provide charge storage through intercalation-based mechanisms. In terms of computational studies, the charge storage process of MXenes has been studied mainly with respect to their pseudocapacitance. Ando and co-workers \cite{Ando2020_storage_MXene} investigated the capacitive properties of \ce{Ti_2CT_x}(T=F, Cl, O or OH). By studying the electronic coupling and the degree of charge transfer between intercalant and host, a characterisation of cation intercalation under conditions imitating full or partial solvation was performed. It was found that fully hydrated intercalated ions had very little electronic coupling to the host material, while on the other hand cations coordinating directly to the MXene showed noticeable electronic interaction as well as charge transfer. According to the group, the capacitance of MXenes can therefore take place on a spectrum between double layer and Faradic mechanisms.

Zhan and co-workers \cite{Zhan2018_understanding_MXene_pseudocapacitance} developed a computational methodology to simulate the capacitive voltage window of MXenes, taking both Faradic and double layer capacitance into account. The results for the prototypical \ce{Ti_3C_2T_x} showed that Faradic pseudocapacitance dominates the capacitive response of the electrode. Double layer capacitance was also observed, but due to the positive surface charge, this process was working against the Faradic capacitance. In a later paper by the group \cite{Zhan2019_screening_MXenes}, the developed method was used to screen for MXenes with desired capacitive properties, including high gravimetric and/or areal capacitance over a large voltage window. From their results, the researchers also attempted to pin down descriptors that relate to these properties. Goff et al. \cite{Goff2021_predicting_window_MXenes} also screened a set of MXenes in search of candidates with a wide capacitive voltage window. For this study, a voltage dependent cluster expansion model based on joint DFT was used. Following their screening, the group concluded that MXenes based on group-VI transition metals showed high voltage windows compared to the more commonly studied Ti-based MXenes.

\subsection{Zn-ion batteries}

    To date, the development of aqueous Zn-ion batteries (ZIBs) is still in its early stage, and many challenges need to be overcome for practical applications. \cite{fang2018recent,wan_energy_2020}  Firstly, elucidating the complicated and controversial electrochemical reaction mechanisms is a difficult task, which hinders researchers from gaining a deep understanding of ZIBs. Secondly, even though the generation of Zn dendrite can be greatly inhibited in the mild electrolytes, but the consumption of Zn metal anode by parasitic side reactions, notably hydrogen evolution reaction (HER), are inevitable and leads to poor reversibility. Thirdly, suitable cathodes materials with high stability in mild acidic electrolytes and high capacity of ion storage are still limited. Theoretical modeling of the aqueous ZIBs will not only provide new insights into the reaction mechanisms of interest, but also accelerate the developments of ZIBs.  \par

    	\subsubsection{Cathode materials in Zn-ion batteries}
	
	Similar to LIBs, the \ce{Zn^2+} will be inserted into or extracted from the cathode material in the working processes of ZIBs. Therefore, the final cost, rate performance and gravimetric energy density are mainly determined by the cathode materials. An ideal cathode should possesses a large storage capacity, a high working voltage, and a stable crystal structure in mild electrolytes. \cite{jia2020active} Even though lots of efforts have been made to design new cathode materials, only limited feasible candidates have been explored, such as manganese (Mn)-based oxides, vanadium (V)-based materials, Prussian blue analogs, carbon-based materials, organic compounds, metal chalcogenides, and MXenes-based materials. For the design and development of new cathode materials, diffusion kinetics of  \ce{Zn^2+} ions also becomes a crucial factor. The ionic radius of the Zn ion is comparable to that of the Li ion (Zn ion: 0.74 Å vs Li ion: 0.76 Å, both at a coordination number of 6), but the large charge density induces strong electrostatic interaction between the host structure and the divalent working ion, leading to sluggish hopping dynamics of the charge carrier, which may diminish the rate capability of ZIBs. \cite{deng_exploring_2023} \par
 
    Several optimization strategies for high-performance manganese dioxide-based ZIBs materials with different crystal forms, nanostructures, morphologies, and compositions are discussed by Jia et al, \cite{jia_recent_2024} which have long been introduced as the cathode materials in ZIBs. Recently, Wang et al.\cite{wang_atomically_2023} reported 2D \ce{MnO_2}/\ce{MXene} superlattice structures with multiple active sites and large interlayer distances where DFT calculations reveal that the regularly layered structure has reduced diffusion energy barriers between \ce{H+/Zn^2+} and host structures leading to highly reversible \ce{H+} and \ce{Zn^2+} insertion/extraction, along with improved electrical conductivity and favorable adsorption energy. 
    \par

    Incorporating metal ions or water molecules into the cathode material layer can help mitigate the high charge density of divalent \ce{Zn^2+} ions. This addition lowers the activation energy for charge transfer at the electrode interface, thereby diminishing the electrostatic interaction between multivalent \ce{Zn^2+} ions and the host framework. Additionally, introducing oxygen can adjust the interlayer spacing and hydrophilicity of the host framework, reducing the number of \ce{Zn^{2+}-H2O} bonds that need to be broken before intercalation. This reduction in bond-breaking significantly decreases the desolvation energy cost and enhances intercalation kinetics. \cite{jia_recent_2024}  In this direction, Shin et al. \cite{shin_hydrated_2019} highlighted the importance of the "hydrated intercalation" mechanism where water molecules are cointercalated with \ce{Zn^2+} ions upon discharge within the lattice framework of \ce{V6O13}. Their experimental techniques and DFT calculations led to two key observations. Firstly, \ce{V6O13} inherently exhibits a single-phase reaction mechanism, allowing it to accommodate ions through reversible expansion and contraction without forming phase boundaries. Secondly, the presence of water enhances Zn (de)intercalation, with the "water effect" contributing to the stable nature of the Zn-intercalated \ce{V6O13} framework in aqueous environments. Water molecules act as coordinating ligands, maintaining the original structural motifs and shielding electrostatic interactions with the host matrix. This results in a low kinetic energy barrier (0.87 eV) (Figure \ref{fig_ZIB}b) for the diffusion of \ce{Zn^2+} ion with water compared to 2.85 eV with water, and subsequently high rate performance can be achieved. \par
    
    Deng and Sun \cite{deng_exploring_2023} proposed two new vanadium-based $\mathrm{ZnVO_3}$ polymorphs ($P2/c-$ and $P2_1/c-$) as cathode candidates for ZIBs. Initial structure search was carried out using the evolutionary algorithm, and ab initio calculations were performed on these new phases to verify the dynamic, mechanical, and thermodynamic stability. Probing electronic conductivity and electrochemical activity revealed the $P2_1/c$-structured $\mathrm{ZnVO_3}$ exhibits a small bandgap of 1.62 eV, a high voltage plateau of 1.16 V, and a trivial volume change upon charging.  Further, CI-NEB results showed low \ce{Zn^{2+}} migration barriers of 0.44 and 0.40 eV for the predicted polymorphs enabling faster diffusion kinetics. Similarly many V-based materials have attracted a growing interest further owing to their improved energy density attributable to multiple valences (ranging from +2 to +5) and delivery of decent cycle lives due to relatively stable structures. \par
    
    Organic cathodes based on p-chloranil employing nanoconfinement strategy has also been reported by Kundu et al. \cite{kundu_organic_2018} where DFT calculations prominently showed that the molecular columns in p-chloranil undergoes a unique twisted rotation to accommodate incoming \ce{Zn^2+} ions as seen in Figure \ref{fig_ZIB}a, thereby minimizing the volume change (-2.7 \%) during \ce{Zn^2+} cycling. Using the RL framework, Fujita and co-authors \cite{fujita2022understanding} trained a RNN-based molecular generator which can generate a series of quinone derivatives. Although the target property is absorption wavelength, this generative model will provide more organic candidates for cathode materials. \par

    High-throughput screening is a promising approach to explore possible excellent cathode materials. Zhou et al \cite{zhou2021machine} screened about 70 new promising cathode materials with high capacity ( 100 mAh $\mathrm{g^{-1}}$) and high voltage (0.5 V) by crystal graph convolutional NN model from over 130 000 inorganic structures.  Combining ML-based screening and AIMD simulations, Li group \cite{cai2021machine} recently reported five new spinel structures with high ionic conductivity ( $\mathrm{10^{-4}}$ $\mathrm{S\cdot {cm}^{-1}}$), which are possible as high-performance cathode materials. 
     Lack of appropriate cathode materials have essentially become a barrier for commercialization of AZIBs. Thus, these studies mainly lays groundwork for employing theoretical calculations to accelerate the discovery of cathode materials.\par

	\subsubsection{Anode materials in Zn-ion batteries}
	
	Due to its moderate redox potential (-0.76 V vs standard hydrogen electrode), high theoretical capacity (5855 mAh $\mathrm{{cm}^{-1}}$), and low polarizability in neutral or mildly acidic aqueous electrolytes, the metallic Zn has been directly used as anode in ZIBs, where the Zn anode undergoes reversible $\ce{Zn/Zn^{2+}}$ electroplating/stripping. Atomic scale modeling for Zn metal is expected to help us better understanding its unique properties. Based on the features created by modified embedded atom method, Nitol and co-workers\cite{nitol2021artificial} trained an ANN interatomic potential for pure Zn, which successfully identified the hexagonally close packed structure as the lowest energy state and predicted the correct c/a ratio by MD simulations. Recently, Mei et al \cite{mei2024development} also developed a ML interatomic potential for the pure Zn system via the moment tensor potential (MTP) framework on a large configuration dataset at DFT level. In addition to the c/a ratio, elastic and defect properties of Zn metal can be accurately predicted by it. The generation of ZnO passivation layer is a drawback for ZIBs, especially in alkaline electrolyte. Behler and co-authors \cite{quaranta2017proton} employed the high-dimensional neural network potential (HDNNP)-based MD simulations to study the dynamics of water-ZnO interface, and the water dissociation/recombination processes via proton transfer can be clearly observed.\par
	
\begin{figure}
        \centering
        \includegraphics[width=0.85\textwidth]{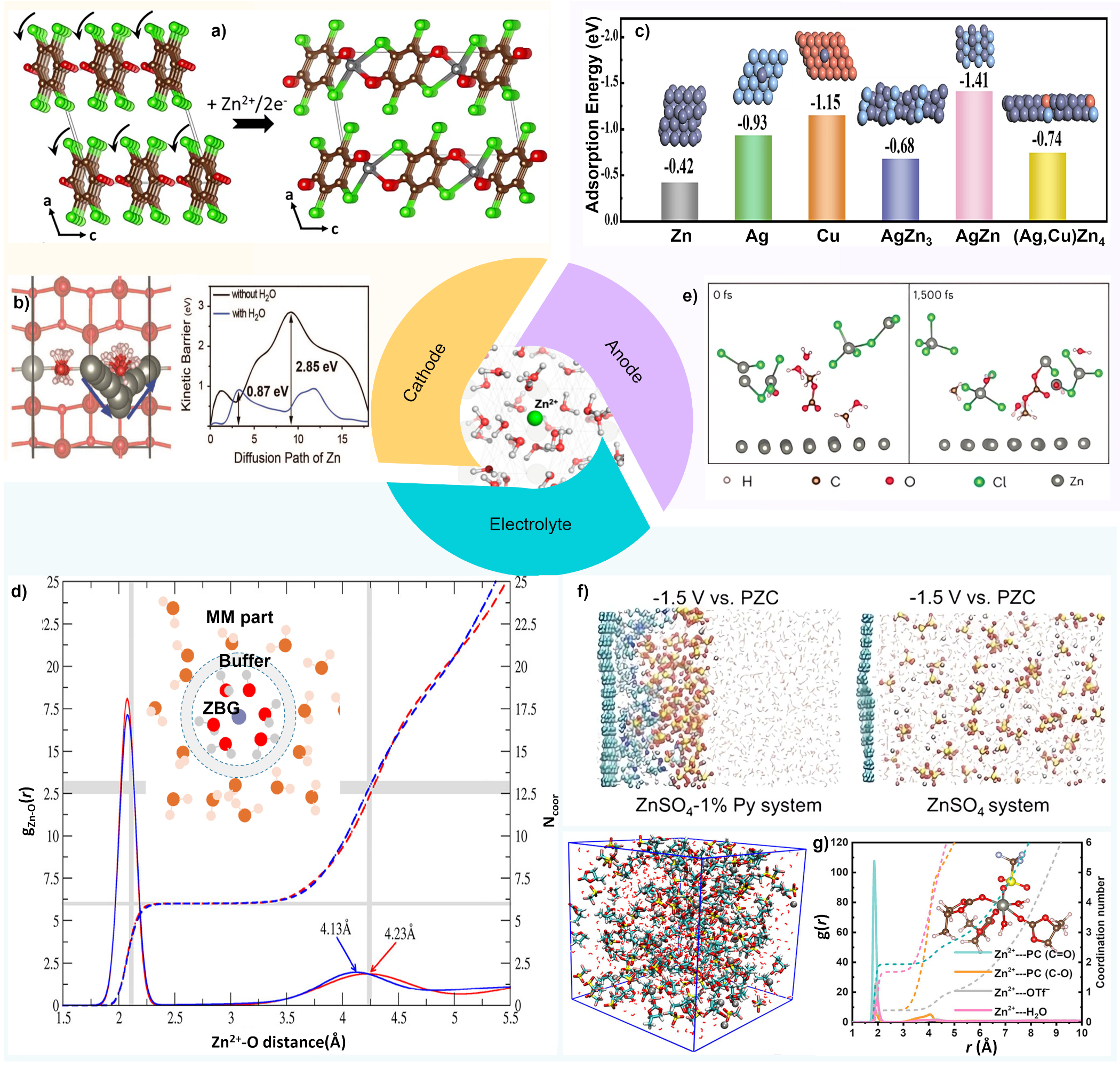}
        \caption{(a) The rotation of the p-chloranil molecular columns upon \ce{Zn^2+} insertion, adapted with permission from Ref.\cite{kundu_organic_2018} (Copyright 2018 American Chemical Society). (b) Diffusion paths of Zn ion with water (projected down [001]) and calculated diffusion barriers for Zn ion with/without water, adapted with permission from Ref.\cite{shin_hydrated_2019} (Copyright 2019 WILEY-VCH Verlag GmbH Co. KGaA) (c) Adsorption energies of \ce{Zn} atoms on \ce{Zn}, \ce{Ag}, \ce{Cu}, \ce{AgZn_3}, \ce{AgZn}, and (\ce{Ag}, \ce{Cu})\ce{Zn_4} substrates, adapted with permission from Ref. \cite{liu_cu-ag_2024} (Copyright 2024 Elsevier Inc.) (d) Radial distribution functions (solid lines) and coordination number (dashed lines) of \ce{Zn^2+} in water system simulated by NN/MM strategy in which the ZBG region was described by NN potential and two charge schemes were applied on water molecules (RESP: red; TIP3P: blue), adapted with permission from Ref. \cite{xu2019molecular} (Copyright 2019 American Chemical Society).   (e) DFTMD simulation snapshot of the formation process of organic SEI species on the \ce{Zn} metal surface at 0 and 1.5 ps, adapted with permission from Ref. \cite{jiang_chloride_2023} (Copyright 2023 Jiang et al.) (f) MD simulation snapshot of \ce{ZnSO4}-1\% Py and \ce{ZnSO_4} with graphene electrode under -1.5 V vs. PZC respectively, adapted with permission from Ref.\cite{luo_regulating_2023} (Copyright 2023 Wiley-VCH GmbH). (g) MD simulation cell snapshot and corresponding RDF plots for 1 M \ce{Zn(OTf)_2} in 50\% PC-sat. solution (2.14 M), adapted with permission from Ref.\cite{ming_co-solvent_2022}(Copyright 2022 American Chemical Society).  }
        \label{fig_ZIB}
    \end{figure}

    Despite the mentioned advantages of Zn metal anode, the unavoidable generation of Zn dendrite, surface corrosion and HER somewhat damage the electrochemical performance of Zn metal anode.
    Many works focus on constructing new structures of active Zn and modifying the anode surface morphology to cope with the above drawbacks. Hao et al. \cite{hao_-depth_2020} reported on a robust and uniform ZnS artificial solid electrolyte interphase (SEI) layer that forms in situ on the surface of Zn. This layer acts as a physical barrier to suppress Zn corrosion and guides the Zn plating/stripping process underneath the layer, thus preventing dendrite growth. Density Functional Theory (DFT) calculations indicated that bonding interactions between sulfur atoms in ZnS and zinc atoms in the Zn metal alter the charge distribution. This results in an uneven charge distribution at the interface, which not only speeds up \ce{Zn^2+} diffusion at the \ce{ZnS}@\ce{Zn} interface but also improves the adhesion of the ZnS layer to the Zn metal. Another highly efficient artificial interface for stable and reversible Zn metal anodes by Cu-Ag double-layer metal coating is constructed on the Zn anode (Zn@Cu-Ag) where DFT calculations was employed to examine the interactions of Zn atoms on \ce{Zn}, \ce{Ag}, \ce{Cu}, \ce{AgZn_3}, \ce{AgZn}, and \ce{(Ag, Cu) Zn_4} substrates to further explore the regulated effect of the protective layers.\cite{liu_cu-ag_2024}(Figure \ref{fig_ZIB}c)  Further, a novel anode material consisting of a hexagonal 1T-Vanadium diselenide (\ce{1T-VSe2}) film on graphene is developed which induces a horizontally (002)-oriented plate-like Zn crystal deposition morphology instead of randomly oriented grains that prompts the compact Zn deposition prohibiting dendrite formation and ultra-stable cycle life. The MEAM force field-based MD simulations visualize Zn nucleation and morphological evolution at the atomistic level.\cite{li_deposition_2022} Onabuta et al. \cite{onabuta_effect_2022} have studied Zn aggregation behavior during electrodeposition with cationic additive species \ce{Li+}. Experimental methods and theoretical calculations employing density functional theory and kinetic Monte Carlo simulations show hat \ce{Li+} affects the surface diffusion of Zn adatoms and thus changing the nucleation and growth during the initial stage of deposition.  \par

    Notably, hosts including carbon based materials such as a bifunctional cellulose nanowhisker graphene (CNG) membrane assembled with graphene (GN) and cellulose nanowhiskers (CNWs) is reported by Zhang et al.\cite{zhang_ultra-long-life_2021}, which serves dual functions as confirmed by both experimental analysis and MD simulations. Firstly, It acts as a desolvation layer to preclude \ce{H2O} molecules encountering the Zn anode and retards the water-induced corrosion reaction. Secondly, It has negative surface charges which can simultaneously generate a deanionization shock by spreading cations but screening anions to obtain redirected Zn deposition. Thus various works highlight the importance of Zn metal anodes and their plating/stripping reversibilty, and subsequently at its electrochemical performance to push the growth of commercial ZIBs.

	\subsubsection{Electrolytes and interphases in Zn-ion batteries}
	
	Optimizing electrolytes compatible with both anodes and cathodes is an effective strategy to enhance the performance of ZIBs. \cite{yan_insight_2022} The solid-electrolyte interface (SEI) formed between the Zn metal anode and electrolytes plays a crucial role in forming a protective layer to control parasitic side reactions, Zn dendrite growth and surface corrosion. Even though low cost, safety and high ionic conductivity favors aqueous electrolytes, their narrow stable electrochemical window poses a significant barrier to their further applications in high voltage batteries. Many works employing different electrolyte engineering strategy have been described here to overcome challenges associated with various electrolyte aspects of ZIBs. \par

    Solvation Structures around aqueous \ce{Zn^{2+}} ions opens windows for underlying reaction mechanisms incorporated with aqueous electrolytes. Malinowski and Śmiechowski\cite{malinowski_solvent_2022} used AIMD simulations to probe solvent exchange events around aqueous \ce{Zn^{2+}} ions in molecular level details and showed that in addition to the presence of hexacoordinated zinc with fixed first hydration shell waters, the solvent exchange around \ce{Zn^{2+}} follows a dissociative mechanism where a transient penta-coordinated \ce{Zn(H_2O)_5^{2+}} is also present. A NN-based interatomic potential has been developed by Zhang and co-workers \cite{xu2019molecular} to describe \ce{Zn^{2+}} and its neighbour water molecules (ZBG), and further been used to study the hydration process of \ce{Zn^{2+}} through NN/MM MD simulations. The well-reproduced simulated radial distribution functions (Figure \ref{fig_ZIB}d) and X-ray absorption near edge structure spectrum confirms the high accuracy of this interatomic potential which may further be employed to find the best operating temperature and salt concentration of target aqueous electrolytes.\par
    
     Li et al. \cite{li_zn_2024}devised a reactive force field for Zn electrode and \ce{ZnSO_4-H_2O} electrolyte and run a reactive MD simulation to understand the influence of pressure on Zn ion diffusion in the electrolyte, the solvation structure of the \ce{ZnSO_4} electrolyte at the interface, Zn reduction kinetics and anode structural evolution at the electrode-electrolyte interface. Adding moderate axial pressure led to the morphology transition from non planar to planar, which is associated with accelerated ion transport and reduction, and thus inhibiting dendrite formation. Added pressure not only induces desolvation of \ce{Zn^{2+}} within the EDL by increase in concentration of unstable \ce{[Zn(H_2O)]^{2+}} or \ce{[Zn(H_2O)_2]^{2+}}, but also improves diffusion to overcome mass transport limitation and thus accelerates reduction kinetics, leading to lower deposition overpotential. \par

    Many organic based electrolyte cosolvents/additives\cite{zhang_electrolyte_2024,zhu_new_2023} have been proposed through examination of  the solvation structure around \ce{Zn^2+} ion using MD simulations. Replacement of some of the \ce{H_2O} molecules in the solvation structure by these additives serves dual role. Firstly, the water activity is greatly reduced, thereby inhibiting the HER and side reactions. Secondly, these additives preferentially adsorbs on the Zn surface to regulate the charge distribution at the Zn anode/electrolyte interface and thus inhibit Zn corrosion and free \ce{H2O}-induced disordered nucleation. Hou et al. \cite{hou_tailoring_2020} reported using acetonitrile (AN) as the electrolyte additive, with overall composition of 1 M \ce{ZnSO4} in hybrid water/acetonitrile mixture, with HWAE-0 denoting 1 M \ce{ZnSO4} in water without the addition of AN and HWAE-X where X denotes volume proportion of acetonitrile. Here first-principles calculations show strong implications on the change of the solvation structure of \ce{Zn^2+} when AN is incorporated into the electrolyte. Further AIMD simulations carried out on HWAE-0 and HWAE-10 suggested that the AN is able to replace up to 3 water molecules in the \ce{Zn(H2O)_6^2+} complex and substantially change the desolvation dynamics of \ce{Zn^2+}, thereby influencing the electrodeposition behavior of Zn metal. \par

    To expand the electrolyte stability window, many water-in-salt (WIS) and water-in-deep eutectic solvent (DES) mixtures has been proposed yet one needs to note that the low cost and high ionic conductivity offered by aqueous electrolyte is hampered. The works of Jiang et al. \cite{jiang_chloride_2023} forms a good example of killing two birds with one stone wherein they not only try to lower the acidity of the electrolyte which suppresses HER, but also stabilize the zinc metal anode by forming an insitu stable SEI layer preventing both HER and dendrite formation. As a consequence, their mixed WIS electrolyte design, 30 m $\ce{ZnCl2}$ + 5 m \ce{LiCl} + 10 m \ce{TMACl} in 5:1 molar ratio of water and DMC, denoted ZLT-DMC attains near-unity CE. They have visualized the formation of an organic SEI layer using AIMD simulations (Figure \ref{fig_ZIB}e) and infer the  importance of LiCl and TMACl for the formation of \ce{[ZnCl]^+} ions whose sufficient concentration is critical for forming a stable and passivating SEI layer.
    
    An interesting work by Zhang et al. \cite{zhang_water_2021} shows the complexity involved with uncovering mechanisms related to solvation of zinc ions. They find results which are fundamentally different from that proposed by Wang et al.\cite{wang_highly_2018}  Wang and coworkers use 1m \ce{Zn-(TFSI)_2} with 20m \ce{LiTFSI} to show increased Zn reversibility and suppressed water activity. They examine changes in \ce{Zn^2+} solvation structures at different concentrations through MD simulations and report that at the highest concentration, \ce{TFSI^-} completely displaces the water molecules about \ce{Zn^2+}, thereby solvating \ce{Zn^2+} with six coordinating oxygen atoms, all from \ce{TFSI^-}. However Zhang and coworkers\cite{zhang_water_2021} revisit the same system and carry out further detailed study using  combined X-ray total scattering, X-ray absorption spectroscopy, FTIR experiments, and classical MD simulations to infer that even at the highest salt concentration of 1m \ce{Zn-(TFSI)_2} and 20m \ce{LiTFSI}, the solvation structure is still \ce{Zn(H2O)_6^{2+}}, and \ce{TFSI^-} anions do not exist to a significant extent in the first solvation shell of  \ce{Zn^2+}. Solving such discrepancies will aid in better understanding of complex interfacial structure and chemistry through both experimental and computational means. \par
          
     Zhao et al. \cite{zhao_water--deep_2019} reported a novel water-in-deep eutectic solvent (water-in-DES) electrolyte, consisting of approximately 30 mol.\% \ce{H2O} in a eutectic mixture of \ce{urea/LiTFSI/Zn(TFSI)2}. Here DFTMD simulations elucidates the structure of electrolyte and shows that all water molecules are confined in the DES matrix due to hydrogen bonding interactions with urea and \ce{TFSI-} and also forming part of \ce{Li+} coordination structure, thereby significantly reducing their reactivity with the Zn anode from both thermodynamic and electrochemical perspectives. Another eutectic 7.6 m \ce{ZnCl_2} aqueous electrolyte with 0.05 m \ce{SnCl_2} additive, which in situ forms a zincophilic/zincophobic \ce{Sn}/\ce{Zn_5(OH)_8Cl_2.H_2O} bilayer interphase where Zincophilic \ce{Sn} decreases \ce{Zn} plating/stripping overpotential and promotes uniform \ce{Zn} plating, while zincophobic \ce{Sn}/\ce{Zn_5(OH)_8Cl_2.H_2O} top-layer suppresses Zn dendrite growth. MD simulations were employed here to investigate the formation of the interfacial structure on the surface of the Zn slab, the nature of the bonding interaction and the hydrogen bonding network of water.\cite{cao_highly_2021} \par

  An electrolyte engineering strategy where a parameter, donor number (DN) of the additive, which estimates the solvation power with \ce{Zn^2+} was chosen to efficiently regulate the Inner Helmholtz plane (IHP) structure with a rational adding amount. Thus, a high DN number pyridine (Py) under small addition (1vol.\%,equivalent to 0.12 M concentration) was proposed by Luo et al. \cite{luo_regulating_2023} where DFT calculations showed that Py exhibited the lowest \ce{Zn^2+}-solvent binding energy and thus steers to reduce \ce{H2O} population within IHP. In Figure \ref{fig_ZIB}f, we can observe the preferential adsorption of Py molecules on graphene electrode surface at -1.5 Z vs. PZC. The effect of polarization voltage on inner Helmholtz plane regions at anode surface through nanosecond MD simulation establishes that the Py molecules form a shielding network between the active anode and the diffusion layer of the electrolyte, passivating the continuous decomposition of bulk free \ce{H2O}. \par

   Zeng et al. \cite{zeng_electrolyte_2021} tackle the challenge of creating a stable SEI layer in situ on a Zn electrode in water-based electrolyte by forming a dense and uniform SEI layer of hopeite \ce{(Zn_3(PO_4)_2.4H_2O)} on the \ce{Zn} surface. This is achieved by simply adding a small amount of \ce{Zn(H_2PO_4)_2} salt to the conventional aqueous electrolyte (1 m\ce{ Zn(CF_3SO_3)_2}) and taking advantage of the local pH increase originating from the competitive side reaction of water decomposition. DFT calculations revealed that the SEI layer manifested a strong adsorption capability toward \ce{Zn^2+} compared that of the bare Zn electrode and a fast \ce{Zn^2+} transport pathway was deduced with lower migration barrier. The hopeite SEI exhibits high interfacial stability, a high Zn-ion transference number, and high Zn-ion conductivity. This not only prevents side reactions by isolating the active Zn from the bulk electrolyte but also ensures uniform and rapid Zn-ion transport kinetics, leading to dendrite-free Zn deposition. \par

     A promising work towards designing a stable and workable anode-free zinc metal batteries (AFZMB) have been proposed by Ming et al. \cite{ming_co-solvent_2022} using the hybrid electrolyte of \ce{Zn(OTf)_2} in PC/water mixture where they assemble an AFZMB by coupling the \ce{Cu} foil with the \ce{ZnMn2O4} cathode leading to extraordinary stability and high capacity retention. This hybrid electrolyte  still inherits the high-safety merit of aqueous electrolytes as their concentrations are much lower than those of common WIS electrolytes. They describe salting-in effect of the addition of \ce{Zn(OTf)_2} in PC/water mixtures. On the basis of experimental characterizations and MD simulations, they observed gradual replacement of  water molecules in the \ce{Zn^2+} primary solvation shell by the \ce{PC} molecules and \ce{OTf^-} anions (Figure \ref{fig_ZIB}g) which significantly reduced water activity and the so-formed hydrophobic SEI served as a protective layer enabling high electrochemical performance and enhanced cathodic stability as well. \par

\subsection{Organic redox flow batteries} 
A typical RFB is integrated by three core components: an electrochemical cell (i.e., two electrodes and a membrane), two outer tanks, and flow pipes. Unlike formed as solid electrodes in conventional batteries, the redox-active materials in RFBs are dissolved into electrolytes.\cite{winsberg2017redox} The catholyte and anolyte are stored in the above two tanks respectively, and separated by the membrane in the electrochemical cell. With this setup, we can recharge the RFBs simultaneously by replacing the discharged electrolytes, as well as scale the battery power and capacity independently. \cite{chen2018recent} These special characters enable the promising large-scale applications of RFBs. \par
\begin{figure}
    \centering
    \includegraphics[width=0.75\textwidth]{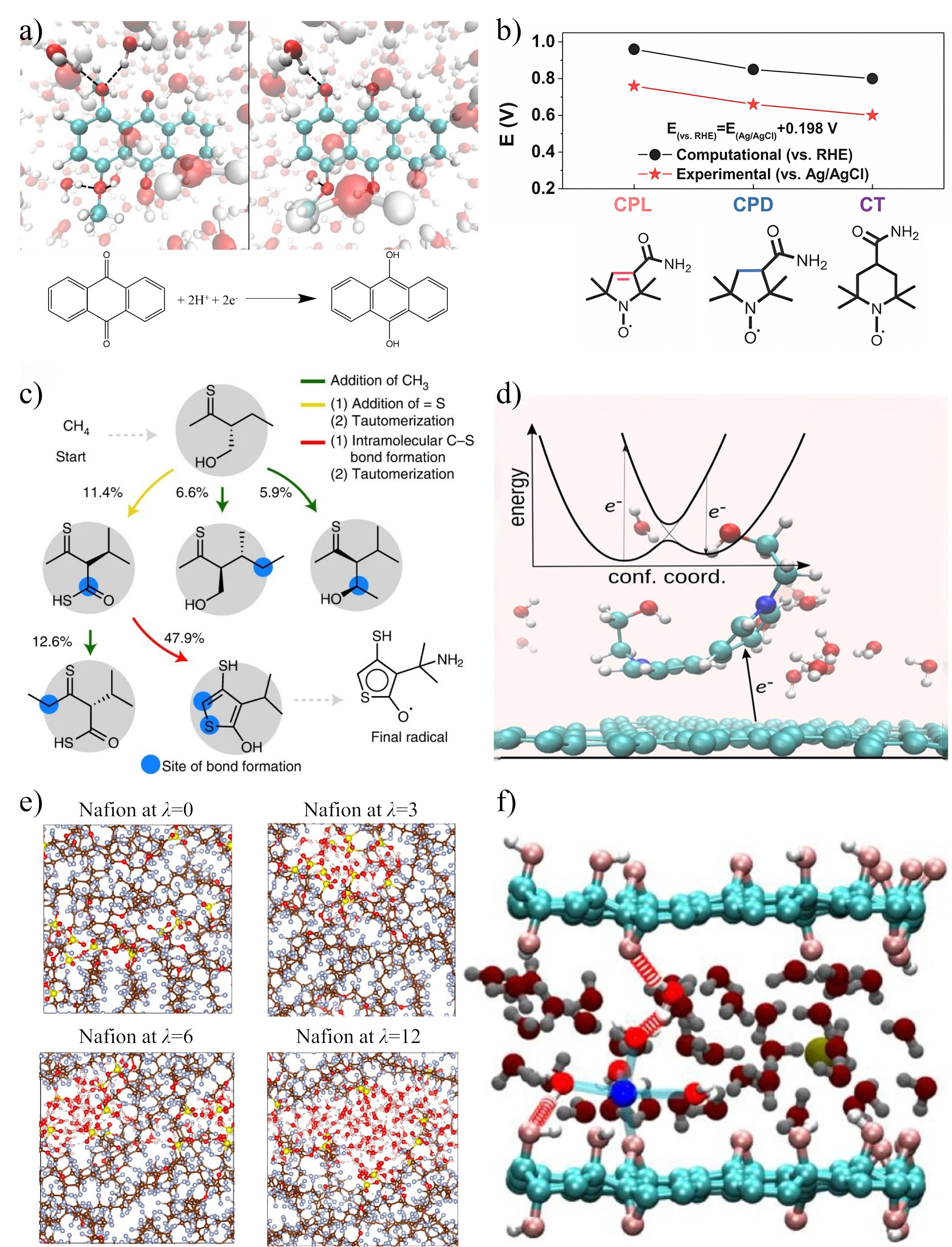}
    \caption{ (a) QM/MM simulations of anthraquinone derivatives, adapted with permission from Ref. \cite{kim2016achieving} (Copyright 2016 American Chemical Society). (b) redox potentials of three cyclic nitroxide radicals, adapted with permission from Ref. \cite{hu2021five}(Copyright 2021 Wiley-VCH GmbH). (c) the RL-based optimization framework for organic radicals, adapted from Ref. \cite{sv2022multi} under the terms of CC-BY licence (Copyright 2022 Sowndarya S.V. et al.). (d) the constrained DFT simulations of the graphene/viologen derivative interface, adapted from Ref. \cite{hashemi2023understanding} under the CC-BY licence (Copyright 2023 Hashemi et al.). (e) SOAP-based MLFF simulation of hydrated Nafion membrane with different water uptakes $\lambda$, adapted with permission from Ref. \cite{jinnouchi2023proton} (Copyright 2023 American Chemical Society). (f) DFTMD simulations of interface between graphene oxide membranes and cationic aqueous solution, adapted with permission from Ref. \cite{chen2017ion} (Copyright 2017 Macmillan Publishers Limited).}
    \label{fig:flowBattery}
\end{figure}

Many modeling and simulation studies have been carried out to investigate the molecular electrochemical properties and to explore promising materials in aqueous organic RFBs (AORFBs). In this section, we will summarize examples in the design of redox-active compounds and membrane materials, as well as the mechanistic elucidations of electrochemical processes in AORFBs with molecular modelling. A summary of these studies is shown in Fig.~\ref{fig:flowBattery}.\par
	
\subsubsection{Redox-active materials}
In general, the redox-active species in catholyte and anolyte are one of the most crucial factor to determine the energy density of AORFBs. There are three aspects should be considered when choosing the active materials.\cite{yang2023organic} Firstly, the redox potential of candidates ought to fall within the electrochemical stability window of aqueous electrolyte, otherwise the water will be decomposed through the hydrogen or oxygen evolution reactions. Secondly, an excellent solubility and chemical stability of the candidates are needed to achieve high energy capacity and long lifespan. Thirdly, the candidates should be abundant and inexpensive to lower the total cost of AORFBs. Due to these stringent requirements, only few types of organic redox-active materials have been employed, such as quinones, nitroxide radicals and aromatic heterocycles. It should be point out that the concept of AORFBs includes two different systems, the one use organic active materials only in a half-cell electrode reaction, while the other use organic active species in both sides of the battery.\cite{winsberg2017redox} \par
Since quinones were reported as redox active materials, related derivatives have received significant attention owing to their strong redox activity, fast reaction rate and excellent electrochemical reversibility. Bachman et al. \cite{bachman2014investigation} studied the influence of R-group substitutions on the redox windows and solvation free energies of about 50 anthraquinone derivatives by DFT calculations. According to their results, a complete methylation of the anthraquinone can increase the reduction window by 0.47 V, while electron-withdrawing R-groups (e.g., chlorine) may decrease the redox window. Using quantum mechanics/molecular mechanics (QM/MM) simulations with explicit water molecules, Zimmerman and co-workers\cite{kim2016achieving} investigated the reduction potentials of 3 special anthraquinone derivatives which can form intermolecular H-bonding with water (Fig.~\ref{fig:flowBattery}a). They found that the theoretical results were in a good correlation with experimental references. After trained a DeepPot-SE model at \textit{ab initio} level as the MLFF, Cheng's group \cite{wang2022automated} estimated the redox potential of OH in water. Hence the hydroxyl groups usually involved into the redox reactions of quinone derivatives, their successful attempts provide a feasible way to accurately predict the reduction potentials of quinone derivatives using high-level MD simulations. Taking the large chemical space of substituents into account, how to efficiently discover and design novel quinone-based active materials is a long-standing challenge. To speed up the discovery, Aspuru-Guzik and co-authors \cite{er2015computational} developed a high-throughput computational screening workflow in which the reduction potentials and solvation free energies of candidate molecules were estimated by DFT calculations. Through a virtual screen process over 1710 quinones/hydroquinones samples, they finally identified about 300 quinones derivatives with a larger predicted reduction potential than 0.7 V. Moreover, a RNN-based molecular generator built by Fujita and colleagues \cite{fujita2022understanding} can generate a series of new quinone derivatives to broaden the diversity of candidate redox-active materials.\par
Radicals are generally considered as an unstable species with high reactivity, whereas some of them can also have an excellent stability under specific steric and/or delocalization effects,\cite{ding2018molecular} for instance, 2,2,6,6-tetramethylpiperidine-1-oxyl (TEMPO). Due to their high redox potentials and fast redox kinetics, these nitroxide-based radicals have been proved as promising redox-active materials in AORFBs.\cite{yang2023organic} The TEMPO is insoluble in pure water, therefore extra supporting electrolyte or modifications with hydrophilic groups are need to enhance its solubility. Pedraza et al.\cite{pedraza2023unprecedented} found that the solubility of TEMPO in lithium bis(trifluoromethanesulfonyl)imide (LiTFSI) aqueous solution is 80 times higher than it in pure water. Further MD simulations revealed that this improvement was achieved by the formation of unique interactions between TEMPO and TFSI anions. Alternatively, Liu and co-authors \cite{liu2016total} increased the solubility of TEMPO by introducing hydroxy groups, and then reported a novel active species 4-hydroxy-2,2,6,6-tetramethylpiperidin-1-oxyl (4-HO-TEMPO) whose solubility in water is up to 2.1 M. Using the primary amide groups, Song and co-workers \cite{hu2021five} synthesized three new cyclic nitroxide radicals based on piperidine, pyrrolidine and pyrroline motifs, respectively (Fig.~\ref{fig:flowBattery}b), among which the pyrroline-based 3-carbamoyl-2,2,5,5-tetramethylpyrroline-1-oxyl (CPL) has the highest redox potential of 0.76 V (vs Ag/AgCl). DFT calculations implied that the high redox potential may steams from its more negative charge density on the N-O radical head and relatively low standard free energy in its reduction reaction. John and colleagues\cite{sv2022multi} developed a molecular optimization framework based on the RL algorithm (Fig.~\ref{fig:flowBattery}c), which can generate target organic radicals with desired redox potential, stability and synthesizability. The properties of top-performing candidates can further be verified by DFT calculations. It is an attractive idea to explore novel nitroxide radicals via this framework, especially using TEMPO as the scaffold.\par
Aromatic heterocycles (e.g., pyrazines, pyrimidines and viologens) is another class of potential redox-active organic molecules because of their high solubility, high stability, and low cost.\cite{yang2023organic} When studying the electrochemical properties of 4-HO-TEMPO, Liu et al.\cite{liu2016total} applied methyl viologen (MV) as the active material in anolyte whose solubility is 3.0 M in 1.5 M NaCl and redox potential is about -0.45 V versus NHE. Amador-Bedolla and co-workers \cite{martinez2020kinetic} analyzed the reduction redox potential and interaction energies of MV species on glassy carbon electrode surface by DFT calculations. They found that the oxidized structure of MV has a lower interaction energy than the reduced one, which is consistent with the adsorptive voltammetric signals. Recently, Hashemi et al.\cite{hashemi2023understanding} investigated the kinetics of electron transfer from graphene electrode to a new viologen derivative 1,1-di(2-ethanol)-4,4-bipyridinium (OH-Vi) by AIMD simulations and the Marcus theory (Fig.~\ref{fig:flowBattery}d). Based on their results, the strong electronic couplings makes the Marcus theory fail, which further lead to a relative big difference between the simulated results and experiments. From this perspective, modeling AORFBs to accurately study the electronic kinetics is still a nontrivial task. Xu and co-authors \cite{liu2020screening} employed DFT calculations to elaborate the structure–performance relationships of hydroxylated viologens. They concluded that redox potentials depends on the energies of lowest unoccupied molecular orbital (LUMO) , while the energy gaps between LUMO and the highest occupied molecular orbital (HOMO) strongly relate to the redox kinetics. Except for viologens, phenazine derivatives also popularly used as the active species. With the help of DFT calculations, Wang and co-workers \cite{hollas2018biomimetic} successfully screened a phenazine-based molecule 7,8-dihydroxyphenazine-2-sulfonic acid (DHPS) which has an excellent solubility (1.8 M in potassium-based supporting electrolyte) and operating voltage (1.4 V vs ferro/ferricyanide catholyte). Also a new derivative 4,4'-(phenazine-2,3-diylbis(oxy))-dibutyric acid (2,3-O-DBAP) recently has been proposed as the active species in anolyte by Kong et al. \cite{kong2023enabling}. Their DFT computational and experimental results shown that this molecule has a low redox potential (-0.699 V vs SHE) and relative greater chemical stability.\par
	
\subsubsection{Membrane materials and membrane–electrolyte interfaces}
The crossover between redox-active species in catholyte and anolyte will lead to a long-term decay of the capacity of AORFBs. \cite{winsberg2017redox} An effective method to minimize this cross-contamination is introducing a membrane to separate two kinds of active materials. It should allows the mobility of positive or negative counter ions on both sides of AORFBs to keep charge neutrality. \cite{ding2018molecular} In addition, a low resistance and a high stability of the membrane are necessary for high performance AORFBs.\cite{winsberg2017redox} In this context, the choice of membrane materials in aqueous electrolyte is still narrow, and there is an urgent to discover new low-cost membranes satisfying the above requirements. \par
So far, cation exchange membranes (CEMs) are the most widely used in AORFBs due to their high ion conductivity and selectivity. Among them, Nafion-based membranes containing hydrophilic sulfonate groups enable rapid proton transfer in aqueous solutions.\cite{chen2018recent} After performed DFT calculations, Sagarik and co-authors \cite{phonyiem2011proton} concluded that at least two structural pathways (i.e., "pass-through" and "pass-by") are possible for the proton diffusion through $\mathrm{-SO}_3\mathrm{H}$ groups in Nafion\textsuperscript{\textregistered} at low hydration levels. Liu and colleagues \cite{liu2023direct} further quantitatively analyzed the surface species during the hydration process of Nafion membrane by ambient-pressure X-ray photoelectron spectroscopy (APXPS) and AIMD simulations. The simulated results were consistent with the APXPS ones, which indicates the high accuracy of AIMD simulations for the membrane–electrolyte interface. To overcome the low efficiency of traditional MD simulations at \textit{ab initio} level, Jinnouchi et al. \cite{jinnouchi2023proton} recently tested a SOAP-based MLFF model through predicting the structure and dynamics of dry and hydrated Nafion (Fig.~\ref{fig:flowBattery}e). Their test results are quite encouraging because not only the hydrophilic and hydrophobic structures but also the proton and water diffusion coefficients had been successfully reproduced. Apart from Nafion, many other CEMs have been proposed in the past decade. For example, by introducing the acidic sulfate ester group to a polybenzimidazole-based compound, Cui's group \cite{pang2021superior} obtained a 1,3-propanediol cyclosulfate grafted polybenzimidazole (OPBI-$\mathrm{OSO}_3\mathrm{H}$) membrane with high proton conductivity. DFT calculations suggested that the proton binding energy of sulfate ester group is lower than it in the sulfonic acid group, so that the former has higher proton dissociation ability and reactive activity. Dai et al. \cite{dai2020thin} fabricated a thin-film composite proton exchange membrane via coating a polyamide-based selective layer on the porous polyethersulfone/sulfonated polyetheretherketone blend (PES/SPEEK) substrate. Both the Grotthuss and vehicle migration mechanism of protons in the polyamide selective layer were observed in further AIMD simulations.\par 
Anion exchange membranes (AEMs) are good selections when cations serving as the redox-active species in RFBs. After selected the poly(ether sulfone) porous membrane as substrate, Yuan and co-workers \cite{hu2021layered} prepared a MgAl-based layered double hydroxides (LDHs) coated composite membrane. The AIMD simulations shown that hydroxide ions can transport through the LDHs layer via both of vehicular and Grotthuss mechanisms. Recently, Sun et al\cite{sun2024anion} proposed a strategy to synthesize anion exchange membranes with orderly channels by self-assembling from electron-rich 1,6-diphenylpyrene and electron-deficient naphthalene diimide-based derivatives. The experimental and DFT computational results demonstrated that these membranes have high conductivity of $\mathrm{Cl}^{-}$ due to the strong electron transfer interactions between different monomers. Chen and colleagues \cite{chen2020effect} investigated the influence of side chain on the electrochemical properties of two poly (ether ether ketone)-based anion exchange membranes via the coarse-grained MD simulations. They emphasized that the ionic conductivity of membrane can be improved by introducing two quaternary ammonium groups in the side chains.\par 
Despite already have many successful applications, the cost of ion exchange membranes accounts for the largest portion of the overall cost of RFBs.\cite{chen2018recent} Using microporous/porous membranes is a promising strategy to decrease the total cost. The electrolyte separation and ion selectivity of these membranes is achieved by the size effect rather than the traditional ion-exchange mechanism. At present, the related researches are meanly focused on microporous polymers and graphenes.\cite{hu2021layered} Based on the hydrophilic polymers of intrinsic microporosity (PIMs), Song and colleagues\cite{tan2020hydrophilic} designed two membranes with high conductivity of salt ions and high size-exclusion selectivity to small organic molecules. The narrow pore size distributions of these microporous membranes were verified by DFT calculations. Xu and co-authors \cite{li2023ultra} reported a series of Tr\"{o}ger’s Base polymers-based microporous membranes which can conduct anion in pH-neutral aqueous electrolytes. They found that the conductivity of these membranes is controlled by the quantity of pores, while the ion transportation energy barrier determined by the pore size distribution. They also preformed MD simulations to study the links between ratios of different monomers and the chain gyration radius of these polymers. A method had been proposed by Chen and co-workers \cite{chen2017ion} to control the interlayer spacing of graphene oxide membranes with \r{a}ngstr\"{o}m precision by introducing a specific cation, and the obtained membranes will selectively exclude other cations with a larger hydrated volumes. DFT calculations and AIMD simulations revealed that the most stable adsorption site is the region where oxidized groups and aromatic rings coexist (Fig.~\ref{fig:flowBattery}f).\par 

\section{Concluding remarks}

In this chapter, we have reviewed both physics-based and machine learning-based techniques used in molecular modelling and their applications to popular aqueous batter systems. This will give a good idea of how it looks like in this rapidly developing field, where there are a lot to be done. To close up, we want to point out the importance of simple and complex when it comes to molecular modelling of (aqueous) batteries. This point is not only useful to young students who just embark on in this field but also relevant as future research directions. 

As J. Willard Gibbs wrote in a letter accepting the Rumford medal of the American Academy of Arts and Sciences in 1881, "One of the principal objects of theoretical research in any department of knowledge is to find a point of view from which the subject appears in its greatest simplicity".~\cite{Zwanzig2001-cl} In this regard, the molecular modelling is the means not the goal. To develop a simple (not simpler) and unified theoretical view (framework) by joining experiment and modelling is one of the missions instead. This requires theoreticians to better understand experiments and to extract the common principles from a plethora of seemingly different experimental observations. Keeping that in mind, theory and modelling community will thrive and play an even more important role in the quest of high-energy density aqueous batteries and related electrochemical devices. 

The other aspect on the spectrum comes from the famous paper ``more is different" written by the condensed matter theorist and Nobel prize laureate Philip W. Anderson~\cite{1972.Anderson}. Electrochemical systems such aqueous batteries is astonishingly complex in disguise of simple operating principles. Therefore, it is inevitable that data-driven approaches such as generative AI will become crucial in the toolbox of molecular modellers to tame the complexity. New computational techniques originated from computer science and their marriages with existing molecular modelling methods will for sure bring the momentum and opportunities to the theoretical developments (of tools) and investigations of aqueous battery systems. 

Aiming at the simplicity and embracing the complexity,  the gap between theory and experiment will be further narrowed down and we will achieve the goal to understand, control and design electrochemical systems at atomistic precision in the end.

\begin{acknowledgement}
This project has received funding from the European Research Council (ERC) under the European Union's Horizon 2020 research and innovation programme (grant agreement No. 949012). This work was partially supported by the Wallenberg Initiative Materials Science for Sustainability (WISE) funded by the Knut and Alice Wallenberg Foundation (KAW). The authors are thankful for the funding from the Swedish National Strategic e-Science program eSSENCE, STandUP for Energy and BASE (Batteries Sweden). 
\end{acknowledgement}

\providecommand{\latin}[1]{#1}
\makeatletter
\providecommand{\doi}
  {\begingroup\let\do\@makeother\dospecials
  \catcode`\{=1 \catcode`\}=2 \doi@aux}
\providecommand{\doi@aux}[1]{\endgroup\texttt{#1}}
\makeatother
\providecommand*\mcitethebibliography{\thebibliography}
\csname @ifundefined\endcsname{endmcitethebibliography}
  {\let\endmcitethebibliography\endthebibliography}{}

\end{document}